\documentclass[aps, amsfonts, nofootinbib, showpacs, byrevtex, preprintnumbers,
showkeys]{revtex4}   

\usepackage{longtable}

\newcommand{\tr}{\text{tr}}
\newcommand{\Tr}{\text{Tr}}
\newcommand{\PP}{\mathsf{P}}
\newcommand{\first}{\right|_{1^{\text{st}}}  \hspace*{-0.7em}}
\newcommand{\firstt}{\right|_{1^{\text{st}}}}
\newcommand{\second}{\right|_{2^{\text{nd}}} \hspace*{-0.7em}}
\newcommand{\secondt}{\right|_{2^{\text{nd}}}}

\newcommand{\be}{\begin{equation}}
\newcommand{\ee}{\end{equation}}
\newcommand{\bea}{\begin{eqnarray}}
\newcommand{\eea}{\end{eqnarray}}

\begin{document}
%
\title{Kamlah Expansion and Gauge Theories}\thanks{supported by DFG
under grants DFG Mu 705/3, DFG III GK-GRK 683/1-01, DFG-Re 856/4-1
and DFG-Re 856/5-1.} 
\date{July 9, 2002}
\author{Oliver Schr\"oder}
\email[Email:]{schroedo@tphys.physik.uni-tuebingen.de}
\affiliation{Eberhard-Karls Universit\"at T\"ubingen}
\author{Hugo Reinhardt}
\email[Email:]{hugo.reinhardt@uni-tuebingen.de}
\affiliation{Eberhard-Karls Universit\"at T\"ubingen}
\preprint{UNITU-THEP-20/02}
%
%
%
\begin{abstract}
In Yang-Mills theories, variational calculations of the Rayleigh-Ritz
type face the problem that on the one hand,  
calculability puts severe constraints on the space of test wave
functionals; on the other hand, the test wave functionals have to be
gauge invariant. The conflict between the two requirements can be
resolved by introducing a projector. In this paper we present an
approach to approximating the projector in a way known and
successfully employed in nuclear physics: the Kamlah expansion. We
discuss it both for electrodynamics and for Yang-Mills theories to
leading order in a perturbative expansion, and demonstrate that the results
are compatible with what one would expect from perturbation theory.
\end{abstract}
%
\maketitle
%
\section{Introduction and Motivation}
Recently, there has been a renewed interest in the application of
variational techniques to Yang-Mills theories, cf.
\cite{Kerman:1989kb,Kogan:1995wf,Diakonov:1998ir,Heinemann:1998cx,
Heinemann:1999ja} and references therein. Whereas the first
investigation in this line \cite{Kerman:1989kb} did not focus
explicitly on the subject of gauge invariance, subsequent
investigations have put more stress upon the 
subject. The basic idea can already been found in the nuclear physics
environment \cite{Ring:1980}: if one starts from a trial wave
functional that does not share the basic symmetries of the
Hamiltonian, one should not vary the expectation value of the energy
but rather introduce a projection operator $\PP$ that projects the
wave functional onto the symmetric space and vary this new expectation
value: 
\be
\delta \frac{\langle \psi | H \PP | \psi \rangle}{\langle \psi | \PP
| \psi \rangle} \stackrel{!}{=} 0.
\ee
This construction allows the usage of simple wave
functionals. It turns out, unfortunately, that the projector acting on
e.g. the simple class of Gaussian wave functionals cannot be evaluated
exactly in the case of Yang-Mills theories. Therefore a number of
approximation methods have been devised. One class of approximations
treats the projection integral as an effective non-linear sigma model
(with a non-local action) and uses common field theoretical
approximation methods to deal with this model \cite{Kogan:1995wf},
cf. also \cite{Diakonov:1998ir}. 
Another approach has suggested an 'improved 
energy functional' \cite{Heinemann:1998cx, Heinemann:1999ja}. The idea
that is followed in this approach is that the presence of the
projector could be simulated by considering an improved energy
functional which contains besides the energy expectation value
additional terms which depend on the generator of the symmetry
$\Gamma^a(\mathbf{x})$: 
\be
E^{eff} = \langle \psi | H | \psi \rangle -
(\Theta^{-1})^{ab}(\mathbf{x}, \mathbf{y}) \langle \psi | \Gamma^a(\mathbf{x})
\Gamma^b(\mathbf{y}) | \psi \rangle
\ee
where we have assumed that the states are properly normalized.
Here $\Theta^{ab}$ denotes the ``moment of inertia'' corresponding to
the ``rotations'' generated by $\Gamma^a(\mathbf{x})$. 
In this paper we want to investigate how one can derive such an
improved energy functional and under which restrictions this can
actually be done. \\
The structure of this paper is as follows: first, in section \ref{sec_proj}, we
present the projector onto gauge invariant states as it is be used in this
paper; we also comment shortly on how one can write down such a
projector in the presence of external colour charges. In section
\ref{sec_kam} we introduce the general concept of the Kamlah expansion as it is
used in nuclear physics \cite{Kamlah:1968, Ring:1980, Blaizot:1986}
, and formulate it in a way s.t. it may be used
also for gauge theories. In section \ref{sec_QED} we apply the
concept to the theory of a quantized electromagnetic field in the presence
of classical charges. There we will see that the leading order of the
expansion - which was often sufficient in the case of nuclear physics -
is in fact inconsistent in quantum field theories. In section
\ref{sec_YMT} we discuss the application to Yang-Mills
theories. However, we are only able to discuss the leading order of
a subsequent perturbative expansion. Also, some additional
difficulties w.r.t. the usage of external charges will become
apparent. In section 
\ref{sec_sum} we give a summary and a critical evaluation of the
Kamlah expansion. Conventions and the canonical quantization scheme
used in this paper are discussed in appendix \ref{app_conv},
some results from perturbation theory will be presented in
appendix \ref{app_PT} and details of some of the calculations are given in
appendix \ref{app_deriv}.

\section{\label{sec_proj}The Projector onto Gauge Invariant States}
\subsection{Generalities}
In this section we want to introduce the concept of the projector onto
gauge invariant states. The basic idea is simple: if we start from a
wave functional $\psi_{\text{ni}}[\mathbf{A}] = \langle \mathbf{A} |
\psi_{\text{ni}} \rangle$  which is not gauge invariant, we can
construct a gauge invariant wave functional $\psi_{\text{gi}}[\mathbf{A}]$ by 
summing over all gauge transformed wave functionals with the correct
Haar measure. Symbolically
\be
\psi_{\text{gi}}[\mathbf{A}] = \int \mathcal{D} \mu[\phi]\,
\psi_{\text{ni}}[\mathbf{A}^U] \label{eq_proj0}
\ee
where $\mathcal{D} \mu[\phi]$ symbolizes the integration over all (as we will
see below: topologically trivial) gauge transformations parametrized
by $\phi$. For notation cf. also app. \ref{app_conv}. For the purposes
of this paper - as outlined in app. \ref{App_canon_quant} - it is
sufficient to have wave functionals that are annihilated by the Gauss
law operator. This is equivalent to having wave functionals that are
invariant under topologically trivial gauge transformations; therefore
the domain of integration in eq.\,(\ref{eq_proj0}) can be restricted
to topologically trivial gauge transformations also. In this case the
gauge transformation can be written with the help of the Gauss law
operator $\mathcal{G}$ (in absence of external charges given in
eq.\,(\ref{eq_def_gamma}), in the presence of external charges a
further discussion can be found in
sec.\,\ref{sec_proj_ext_charges}):  
\be
 \psi_{\text{ni}}[\mathbf{A}^U] = e^{- i \int d^3x\, \phi^a(\mathbf{x})
 \mathcal{G}^a(\mathbf{x})}  \psi_{\text{ni}}[\mathbf{A}]. 
\ee
This now allows to rewrite  eq.\,(\ref{eq_proj0})  in terms of an
operator $\PP^{\rho}$ acting upon $\psi_{\text{ni}}$: 
\be
\psi_{\text{gi}}[\mathbf{A}] = \left( \int \mathcal{D} \mu[\phi] e^{-
i \int d^3x\, \phi^a(\mathbf{x}) 
\mathcal{G}^a(\mathbf{x})} \right) \psi_{\text{ni}}[\mathbf{A}] = \PP^{\rho}
\psi_{\text{ni}}[\mathbf{A}] \label{eq_proj1},
\ee
where $\mathcal{D} \mu[\phi]$ denotes the integral over all variables
parametrizing the gauge transformation including the Haar measure
necessary to make the integral invariant under left multiplication.

\subsection{Projector in the presence of external charges\label{sec_proj_ext_charges}}
If we restrict ourselves to the pure gauge sector without external
charges the Gauss law
operator is simply (for conventions and placement of factors of g
cf. appendix \ref{app_conv}, $\Gamma^a$ is defined in
eq.\,(\ref{eq_Gauss_law_constraint_puregauge}))  
\be
\mathcal{G}^a(\mathbf{x}) = - \Gamma^a(\mathbf{x}) 
\label{eq_def_gamma} 
\ee
Both the Gauss law operator and its pure gauge part fulfil the
``angular momentum'' algebra 
\be
[\mathcal{G}^a(\mathbf{x}), \mathcal{G}^b(\mathbf{y})] = i f^{abc} {\cal
G}^c(\mathbf{x}) \delta_{\mathbf{x} \mathbf{y}}, \hspace*{3em} 
[- \Gamma^a(\mathbf{x}), -\Gamma^b(\mathbf{y})] = i f^{abc}
(-\Gamma^c(\mathbf{x}))  \delta_{\mathbf{x} \mathbf{y}}.
\label{eq_comm_rel_Gauss_law}
\ee
If we want also to consider external charge contributions, this can be
done most economically in the following way (for definiteness, we
assume the external charges to be fermions, in the following
generically called 'quarks') \cite{Polyakov:1987ez, Zarembo:1998qm,
Zarembo:1998xq}: add to the Hamiltonian of the gauge
sector a matter field part $H_M$:
\be
H_M = M \int d^3x\,\bar{\psi}(\mathbf{x}) \psi(\mathbf{x})
\ee
where $M$ denotes the quark mass, and we treat a single quark species
(``flavour'')  for simplicity.
In concordance with the common interpretation we associate
\textit{creation} operators with operators that create positive energy
states, and thus write explicitly
\be
\psi^{a}_{\alpha}(\mathbf{x}) = \left(\begin{array}{c} a^a_1(\mathbf{x}) \\
a^a_2(\mathbf{x}) \\ 
b^{\dagger a}_1(\mathbf{x}) \\ b^{\dagger a}_2(\mathbf{x}) \end{array} \right)
\hspace*{0.25em} , \hspace*{1.75em} \psi^{\dagger a}_{\alpha}(\mathbf{x}) =
\left(\begin{array}{c} a^{\dagger a}_1(\mathbf{x}) \\ a^{\dagger a}_2(\mathbf{x}) \\
b^{a}_1(\mathbf{x}) \\ b^{a}_2(\mathbf{x}) \end{array} \right). 
\ee
We then interpret the operators $a, a^{\dagger}$ as connected to
quarks, and the operators $b, b^{\dagger}$ as connected to anti-quarks. 
We now turn to the charge
operator. Let $\lambda^a$ be the (hermitian) generators of the SU(N)
representation appropriate for the quarks. The charge operator is then
given by 
\be
\rho^a(\mathbf{x}) = J_0^a(\mathbf{x}) = \bar{\psi}(\mathbf{x}) \gamma^0
\lambda^a \psi(\mathbf{x}), 
\ee
where there is no integration over $\mathbf{x}$ just as in the remainder
of this section (unless explicitly stated).
In terms of the $a,b$ operators, it can be written as 
\be
\rho^a(\mathbf{x}) = a^{\dagger b}_i(\mathbf{x}) \lambda^{a}_{bc}
a^{c}_i(\mathbf{x}) - b^{\dagger b}_i(\mathbf{x}) \lambda^{a*}_{bc}
b^{c}_i(\mathbf{x}).  
\ee
The matrix $\lambda^*$ denotes simply the complex conjugate matrix of
$\lambda$. The charge operators  
inherit their commutation relations from the $\lambda$ matrices:
\be
~[\rho^a(\mathbf{x}), \rho^b(\mathbf{y})] = \delta_{\mathbf{x}\mathbf{y}} \left(
a^{\dagger}_i(\mathbf{x}) [\lambda^a, 
\lambda^b ] a_i(\mathbf{x}) + b^{\dagger}_i(\mathbf{x}) [\lambda^{a*},
\lambda^{b*} ] b_i(\mathbf{x}) \right) = i f^{abc} \rho^c(\mathbf{x})
\delta_{\mathbf{x}\mathbf{y}}. 
\ee
Thus, we can form  the Gauss law operator in the presence of external
charges
\be
\mathcal{G}^a(\mathbf{x}) = - \Gamma^a(\mathbf{x}) + \rho^a(\mathbf{x}),
\label{eq_def_gamma_plus_rho}
\ee 
where $\Gamma^a(\mathbf{x})$ is defined in
eq.\,(\ref{eq_Gauss_law_constraint_puregauge}).  
We form states in the presence of external charges (called $|
\text{coupled} \rangle$)  in the following 
way:
\be
| \text{coupled} \rangle = \sum_{a,i} c^a_i | \text{fermion}~ i \rangle |
  \text{YM} ~a \rangle, \label{def_coupled_states}
\ee 
where $c^a_i$ are probability amplitudes and $| \text{YM} ~a \rangle$
are purely gluonic states labelled by $a$. The fermionic states are 
generated by acting 
with quark and anti-quark creation operators on the (gauge invariant)
fermionic vacuum \footnote{That the fermionic vacuum is gauge invariant
is a trivial consequence of the fact that it is annihilated by the
annihilation operators.}:
\be
| \text{fermion}~ i \rangle = a^{\dagger}_{i_1} \cdots
  a^{\dagger}_{i_n} b^{\dagger}_{j_1} \cdots b^{\dagger}_{j_m} | 0
  \rangle, \label{creation_of_fermionic_states}
\ee
where we have combined colour-, spinor- and spatial indices into a single
super-index $i$ (or $j$), and denoted the fermionic vacuum  by
$|0\rangle$. We can also write suggestively
\be
\sum_{a,i_1,\ldots, j_m} c^a_{i_1 \ldots j_m} a^{\dagger}_{i_1} \cdots
  a^{\dagger}_{i_n} b^{\dagger}_{j_1} \cdots b^{\dagger}_{j_m} | 0
  \rangle | \text{YM} ~a \rangle\, = 
\sum_{i_1,\ldots, j_m}
  a^{\dagger}_{i_1} \cdots a^{\dagger}_{i_n} b^{\dagger}_{j_1} \cdots
  b^{\dagger}_{j_m} | 0 \rangle | \text{YM}  \rangle_{i_1 \ldots j_m}
\ee
with $| \text{YM}  \rangle_{i_1 \ldots j_m} = \sum_a  c^a_{i_1 \ldots
j_m} | \text{YM} ~a \rangle .$  Using 
\be 
\begin{array}{ccccc}
e^{-i \int d^3y\, \varphi^b(\mathbf{y}) \rho^b(\mathbf{y})}
\psi_{\alpha}^a(\mathbf{x}) e^{i \int d^3y\, \varphi^b(\mathbf{y})
\rho^b(\mathbf{y})} & = & (e^{i \varphi^b(\mathbf{x}) \lambda^b})_{ac} 
\psi_{\alpha}^c(\mathbf{x}) &=& \mathfrak{q}^{R}_{ac}[U(\mathbf{x})]
\psi_{\alpha}^c(\mathbf{x}), \\   
e^{-i \int d^3y\, \varphi^b(\mathbf{y}) \rho^b(\mathbf{y})}
\psi_{\alpha}^{\dagger a}(\mathbf{x}) 
e^{i \int d^3y\, \varphi^b(\mathbf{y}) \rho^b(\mathbf{y})} & = &
\psi_{\alpha}^{\dagger 
c}(\mathbf{x}) (e^{- i \varphi^b(\mathbf{x}) \lambda^b})_{ca} 
&=& \mathfrak{p}^{R}_{ac}[U(\mathbf{x})]
\psi_{\alpha}^{\dagger c}(\mathbf{x})
\end{array} \label{eq_charge_trafo}
\ee
where $\psi$ transforms according to the representation
$\mathfrak{q}^{R}_{ac}[U(\mathbf{x})]$ and $\psi^{\dagger}$ transforms
according to the conjugate representation
$\mathfrak{p}^{R}_{ac}[U(\mathbf{x})] =
\mathfrak{q}^{R}_{ca}[U^{-1}(\mathbf{x})]$, the requirement  
\be
e^{- i \int d^3x \phi^a(\mathbf{x}) \mathcal{G}^a(\mathbf{x})} |
\text{coupled} \rangle = | \text{coupled} \rangle
\ee
leads to (basically) the same result as known in the literature
\cite{Haagensen:1997pi} where one considers the Yang-Mills wave
functional $| \text{YM} \rangle_{i_1 \ldots j_m}$ with the extra colour
indices:
\be
e^{ i \int d^3x\, \phi^a(\mathbf{x}) \Gamma^a(\mathbf{x})} | \text{YM}
\rangle_{i_1 \ldots j_m} = \mathfrak{q}^R_{i_1 l_1}[U] \cdots
\mathfrak{q}^R_{i_n l_n}[U] \mathfrak{p}^R_{j_1 k_1}[U] \cdots
\mathfrak{p}^R_{j_m k_m}[U] | \text{YM} \rangle_{l_1 \ldots k_m}.
\ee
where the super-index notation for the representation matrices
$\mathfrak{q}, \mathfrak{p}$ has to be
understood as follows: if we decompose the super-index $i_1$ into the
colour index $a_1$, the spatial index $m_1$ and the position index
${\bf x}_1$, and correspondingly $l_1$ into $b_1$, $n_1$ and ${\bf
x}_1$ (the position index is in both cases identical since we do not
associate an integration over position indices with gauge
transformations), then 
\be
(\mathfrak{q}^R,\mathfrak{p}^R)_{i_1 l_1}[U] = \delta_{m_1 n_1}
(\mathfrak{q}^R,\mathfrak{p}^R)_{a_1 b_1}[U({\bf x}_1)]. 
\ee
The formulation chosen in this section has the advantage that one can give
a closed expression for the projector even in the presence of external
charges. At this point one has to note that from the fact that ${\cal
G}^a | \psi \rangle = 0$ one may not conclude that one has
diagonalized all generators $\Gamma^a$ of the gauge part of the Gauss
law operator, since the matter part $\rho^a$ of $\mathcal{G}^a$ is also
operator valued. One can only choose the Cartan subalgebra of the
colour charges to be diagonalized. 
\subsection{The Projector onto Physical States\label{sec_proj_on_phys_states}}
Since we now have an operator that generates gauge transformations on
arbitrarily charged states, we can also write down a very simple
(formal) expression for the projector onto the physical
sector:
\be
\PP^{\rho} = \int \mathcal{D} \mu[\varphi]\, e^{i \int d^3x\,
\varphi^a (\mathbf{x}) (\Gamma^a(\mathbf{x}) -
\rho^a(\mathbf{x}))}. \label{eq_def_gen_proj} 
\ee
The integration domain runs over the topologically
trivial sector of the gauge transformations.  
Two points deserve attention: first, we want to consider why
$\PP^{\rho}$ is indeed a projector onto the physical subspace, and
second, why it is sufficient to restrict the domain of integration to
topologically trivial gauge transformations. In fact these two points
are closely related. A state in the physical subspace is characterized
by the fact that it is annihilated by the Gauss law operator: ${\cal
G}^a | \psi \rangle = 0$. An identical requirement is $e^{-i \int \phi^a
\mathcal{G}^a} | \psi \rangle = | \psi \rangle$, in other words a state
has to be invariant under topologically trivial gauge
transformations. We have abbreviated here $\int d^3x\,
\phi^a(\mathbf{x}) \mathcal{G}^a(\mathbf{x})$ as $\phi^a
\mathcal{G}^a$. If we now act with  $e^{-i \int \phi^a \mathcal{G}^a}$ upon
$\PP^{\rho}$ we can use (a) the fact that topologically trivial gauge
transformations form a group, i.e. 
\be
e^{-i \int \phi^a \mathcal{G}^a} e^{-i \int \varphi^b \mathcal{G}^b} =
e^{-i \int \alpha^a(\phi, \varphi) \mathcal{G}^a},
\ee
(b) the left invariance of the Haar measure
\be
\mathcal{D} \mu[\varphi] = \mathcal{D} \mu[\alpha(\phi, \varphi)],
\ee
and (c) the fact that a topologically trivial gauge transformation does
not change the domain of integration of the projector $\PP^{\rho}$ 
to obtain
\bea 
 e^{-i \int \phi^a \mathcal{G}^a} \PP^{\rho} &=&  e^{-i \int \phi^a
\mathcal{G}^a} \int \mathcal{D} \mu[\varphi]\, e^{-i \int \varphi^b
\mathcal{G}^b} = \int \mathcal{D} \mu[\varphi]\, e^{-i \int \phi^a
\mathcal{G}^a} e^{-i \int \varphi^b {\cal G}^b}  \nonumber \\ &=&  \int
\mathcal{D}\mu[\varphi]\,  e^{-i \int \alpha^a(\phi, \varphi)
\mathcal{G}^a}  =  \int \mathcal{D} \mu[\alpha]\, e^{-i
\int \alpha^a \mathcal{G}^a}  
= \PP^{\rho}. 
\eea
Thus we have answered both questions: the projector projects onto the
physical subspace since the projector is invariant under small gauge
transformations, and since the Gauss law constraint is equivalent to
topologically trivial gauge transformation only this domain of
integration is required.
\section{\label{sec_kam}General Concept of the Kamlah Expansion}
\subsection{Abelian Case}
The concept of the Kamlah expansion \cite{Kamlah:1968} is most easily
explained using an Abelian quantum mechanical example
\cite{Kamlah:1968, Ring:1980, Blaizot:1986}; assume that we have a
two-dimensional system. The generator of rotations is $\hat{J}$. The
projector onto states having angular momentum $I$ is
\be
\PP^I = \int_0^{2 \pi} \frac{d \alpha}{2 \pi} e^{i \alpha (\hat{J} -I)}.
\ee
That this is indeed the correct projector can again be seen most easily by
\be
e^{i \beta \hat{J}} \PP^I = \int_0^{2 \pi} \frac{d \alpha}{2 \pi}
e^{i (\alpha+\beta) \hat{J} - i (\alpha+\beta) I + i \beta I} = e^{i
\beta I} \PP^I   
\ee
where it has been used that $\int_0^{2 \pi} = \int_{\beta}^{2 \pi +
\beta}$ due to the periodicity of the integrand. \\
The case where the Kamlah expansion works best is the case of
\textit{strong deformations}. 
 Strongly deformed states $| \phi \rangle$ are states where
\be
| \langle \phi | e^{i \alpha \hat{J}} | \phi \rangle |
\ee
decreases rapidly for increasing $\alpha$. This is very useful, since
one can argue that
\be
| \langle \phi | H e^{i \alpha \hat{J}} | \phi \rangle |
\ee
will vanish rapidly for increasing $\alpha$, too, but 
\be
\frac{ \langle \phi | H e^{i \alpha \hat{J}} | \phi \rangle }{ \langle
\phi | e^{i \alpha \hat{J}} | \phi \rangle} 
\ee
is a smooth function of $\alpha$.
The argument consists basically of noting that, whereas $\hat{J}$ is a
one-particle operator (in many-body language), $e^{i \alpha \hat{J}}$
is a collective operator that affects arbitrarily large numbers of particles
\be 
e^{i \alpha \hat{J}} = \underbrace{1}_{0-body \  operator} +
\underbrace{i \alpha \hat{J}}_{1-body \ operator} + \underbrace
{ \frac{1}{2} (i \alpha \hat{J})^2}_{2-body \ operator} + \underbrace
{ \frac{1}{3!} (i \alpha \hat{J})^3}_{3-body \ operator} + \ldots \, ,
\ee 
$H$ is (in nuclear physics) at most a two-body
operator (A similar terminology can also be constructed for
Yang-Mills theories, which is done in
\cite{schroedo:2002a, schroedo:2002b}. In this terminology the Yang-Mills
Hamiltonian also will be at most a two-body operator, i.e. contain up
to four creation/annihilation operators.); thus
the behaviour w.r.t. increasing $\alpha$ should be qualitatively the
same whether we 
consider $e^{i \alpha \hat{J}}$ or $He^{i \alpha \hat{J}}$. One 
can now argue further that in the matrix element $\langle \phi | H e^{i
\alpha \hat{J}} | \phi \rangle$ all degrees of freedom that are not
affected by the collective rotation ('internal degrees of freedom')
can be integrated out, and that this results in an effective
Hamiltonian operator that is an - up to now - unknown function of
$\hat{J}$, the symmetry generator. For this function it is proposed to
use a power series expansion in powers of 
$(\hat{J} - \langle \hat{J} \rangle)$ (where $\langle \hat{J}
\rangle$ is used as an abbreviation for $\langle \phi| \hat{J} |\phi
\rangle$), which is plausible,
since one does not expect singularities for collective rotations with
finite angular momenta.
(The idea to  expand the Hamiltonian in
powers of the symmetry generators, although with a different method to
obtain the coefficients, was proposed in
\cite{Villars:1965a, Villars:1965b}, cf. also \cite{MW69, MW70}.)
However, this is only useful if we can stop the expansion
after a few terms. That this is indeed possible one can see as follows: \\
\begin{itemize}
\item by assumption, $\langle \phi | e^{i \alpha \hat{J}} | \phi
\rangle$ is well localized in $\alpha$ space, since we only consider
strongly deformed states; therefore, a broad range of wave functions
with good $\hat{J}$ quantum number have to be added up coherently
\item  $\langle \phi |( \hat{J} - \langle \hat{J} \rangle) e^{i \alpha
    \hat{J}}  | \phi \rangle$: each 
    wave function with good $\hat{J}$ quantum number is now weighted by that
    quantum number and this begins to destroy the coherence, in other
    words 
    $\langle \phi | (\hat{J} - \langle \hat{J} \rangle) e^{i \alpha
    \hat{J}}  | \phi \rangle$ 
    is broader in $\alpha$ space than $\langle \phi |e^{i \alpha
    \hat{J}} | \phi \rangle$ 
\item the higher the power of $(\hat{J} - \langle \hat{J} \rangle)$,
    the broader is the resulting matrix 
  element in $\alpha$ space, \textit{but} we already know that
  $\langle \phi | H e^{i \alpha \hat{J}}  | \phi \rangle$ is not very
  much broader in $\alpha$ space than $\langle \phi |  e^{i \alpha
\hat{J}} | \phi \rangle$, hence the matrix elements containing higher powers of
      $(\hat{J} - \langle \hat{J} \rangle)$ (besides $e^{i \alpha
  \hat{J}}$) must be suppressed (We assume here that there is
  no 'conspiracy' between different orders in the expansion, but this
  seems to be a plausible assumption.), 
      i.e. the corresponding expansion coefficients must be small. qed.
\end{itemize} 
Thus, we can stop the expansion
\be
\langle \phi | H e^{i \alpha \hat{J}} | \phi \rangle = \sum_{k=0}^{\infty}
A_k \langle \phi | (\hat{J} - \langle \hat{J} \rangle)^k e^{i
\alpha \hat{J}} | \phi \rangle 
\label{Abelian_Kamlah_exp1} 
\ee
after a few terms:
\be
\langle \phi | H_{approx} e^{i \alpha \hat{J}} | \phi \rangle =
\sum_{k=0}^{n} A_k \langle \phi | (\hat{J} - \langle \hat{J}
\rangle)^k e^{i \alpha \hat{J}} | \phi \rangle. 
\label{Abelian_Kamlah_exp2} 
\ee
The (n+1) coefficients are determined by requiring
that $\langle \phi | H e^{i \alpha \hat{J}} | \phi
\rangle$ and $\langle \phi | H_{approx} e^{i \alpha \hat{J}} |
\phi \rangle$ along with their
first n derivatives w.r.t. $\alpha$ 
are equal at the point $\alpha=0$. One thus ends up with
the set of equations\footnote{In order to obtain a more compact
notation, we have equated $\langle \phi | H e^{i \alpha (\hat{J} -
\langle  \hat{J} \rangle)} | \phi
\rangle$ and $\langle \phi | H_{approx} e^{i \alpha (\hat{J} - \langle
\hat{J} \rangle)} | \phi \rangle$. In the sections in field theory, we
will equate the terms equivalent to $\langle \phi | H e^{i \alpha (\hat{J} -
I)} | \phi \rangle$ and $\langle \phi | H_{approx} e^{i \alpha
(\hat{J} - I)} | \phi \rangle$. They differ only by a multiplicative
c-number $e^{-i\phi\langle \hat{J} \rangle}$ or $e^{-i\phi I}$
respectively.}
\be
\langle \phi | H (\hat{J} - \langle \hat{J} 
\rangle)^{m} | \phi \rangle =  \sum_{k=0}^{n} A_k \langle \phi |
(\hat{J} - \langle \hat{J} \rangle)^{k+m}| \phi \rangle, 
\label{Abelian_Kamlah_exp3}  
\ee 
where $m=0,\ldots,n$. After we have solved these equations to obtain
$A_i$, we can insert $H_{approx} = \sum_{k=0}^{n} A_k (\hat{J} -
\langle \hat{J} \rangle)^k$ into the expression for the projected
energy \\
\bea
E^{proj}_{approx} (I) &=&  \frac{\langle \phi |
H_{approx} \PP^I | \phi \rangle}{\langle \phi | \PP^I | \phi \rangle}
= \sum_{k=0}^{n} A_k \frac{\langle \phi | (\hat{J} - \langle \hat{J}
\rangle)^{k} \PP^I | \phi \rangle}{\langle \phi | \PP^I | \phi \rangle}. 
\eea
The advantage of the approach now becomes visible: whereas before we did not
know how to evaluate the matrix elements including the projector, it has now
become trivial; $\PP^I$ projects onto angular momentum I. Then
$\hat{J} \PP^I = I \PP^I$,  and numerator and denominator
cancel, which means that we \textit{do not have to calculate a single matrix
  element containing the projector explicitly}:
\be
E^{proj}_{approx}(I) =  \sum_{k=0}^n A_k (I-\langle \hat{J} \rangle)^k.
\ee  
In nuclear physics, for most applications $n$ is taken to be one or
two; one then ends up with
\be
E^{proj}_{approx}(I) = \langle  H   \rangle + \frac{\langle 
H \Delta \hat{J} 
\rangle}{ \langle  (\Delta \hat{J})^2 \rangle} (I- \langle
\hat{J} \rangle) 
\ee
for $n=1$ and
\bea
E^{proj}_{approx}(I) & = & \langle  H   \rangle - \frac{\langle 
 \Delta \hat{J}^2  
\rangle}{2 \mathcal{I}_Y} + \frac{\langle \hat{J} \rangle}{{\cal
I}_{sc}}(I - \langle \hat{J} \rangle) + \frac{1}{2
\mathcal{I}_Y} (I- \langle \hat{J} \rangle)^2,  \\
\text{with}  & &
\frac{1}{2 \mathcal{I}_Y} = \frac{\langle H \Delta \hat{J}^2 \rangle
- \langle H \rangle \langle \Delta \hat{J}^2 \rangle}{\langle \Delta
\hat{J}^4  \rangle - \langle \Delta \hat{J}^2 \rangle^2}~~\text{and}~~
\frac{\langle \hat{J} \rangle }{\mathcal{I}_{sc}} = \frac{\langle H \Delta \hat{J}
\rangle}{ \langle  \Delta \hat{J}^2 \rangle}  \label{mom_of_inertiaYandSC}
\eea
for $n=2$ \cite{Ring:1980}.
We have abbreviated $\langle \phi | \ldots | \phi \rangle$ 
by $\langle  \ldots \rangle$ and used the notation $\Delta \hat{J} =
\hat{J} - \langle \hat{J} \rangle$. 
If we take simply $n=0$, we end up with $A_0 = \langle H \rangle$
which is reassuring: 
the lowest order term just gives the unprojected expectation value.
We see that, at first order, we get no
corrections to the mean field energy if $\langle \hat{J} \rangle = I$,
whereas for the expansion to second order, we even then obtain a
correction. One can interpret this correction if one considers a
deformed state as being a superposition of states with different
angular momenta; 
especially, there are components in the wave function that are of
higher angular momentum than the angular momentum one projects onto. 
Usually one would assume that higher angular momenta also contain
higher kinetic energy that should not be present in the projected wave
function; thus one obtains a negative correction to the ordinary energy
expectation value. One notes in passing that the two moments of
inertia $\mathcal{I}_Y, \mathcal{I}_{sc}$ are quite different from one
another; the latter is under certain circumstances identical to the moment of
inertia of the cranking model \cite{Ring:1980}, since one can show 
that, if the variational space considered for the cranking calculation
contains not only $| \phi \rangle$ but also
$\Delta \hat{J} | \phi \rangle$, then the solution of the cranking
equations indeed solves 
also the variational equations derived for the $n=1$ Kamlah
expansion, and that the Lagrange multiplier of the cranking model has
precisely the same form as the first order correction of the Kamlah
expansion \cite{Mang:1975} \footnote{The variational space that we use
in field theory, consisting of Gaussian wave functionals only, is not so
large as it is required for this equivalence.}. \\ 
There are calculations that implicitly assume that in practice
the moment of inertia $\mathcal{I}_Y$ can be determined by a cranking
type calculation \cite{Heinemann:1998cx, Heinemann:1999ja}.
In Yang-Mills theories it turns out that to $\mathcal{O}(g^0)$ this is
indeed the case, but the general expressions given above in
eq.\,(\ref{mom_of_inertiaYandSC}), and the fact that the variational
space of Gaussian wave functionals is to small to prove the
equivalence of $\mathcal{I}_{sc}$ and the cranking model moment of
inertia,  do not give much hope that it stays this way for higher
orders in perturbation theory.  
\subsection{Non-Abelian Case}
If one goes from an Abelian type problem to a non-Abelian problem (like
three dimensional angular momentum projection) one immediately has to
face some problems \cite{Sorensen:1977}; the first one is that 
a rigid body has two frames of reference
(the lab frame and the intrinsic frame), two sets of angular momentum operators
and (since the total angular momentum is identical in the intrinsic and in the
lab frame) three angular momentum quantum numbers (usually called I,K,M
belonging to $J^2=I^2,J_3,I_3$) \cite{Blaizot:1986}. The basic
ingredient into any projector
is then
\be
\PP^{I}_{KM} = |I, K \rangle \langle I, M|.
\ee
The quantum numbers I,K are those of the state to be observed in the lab
frame, whereas M is only seen in the intrinsic frame. The object
$\PP^{I}_{KM}$ does not have the properties usually expected of a
projector, since it is 
neither hermitian, nor do we have a relation like $\PP^2 = \PP$. The
obvious remedy cannot be used, since taking $\PP^{I}_{K=M,M}$ would
avoid the mathematical problems but would violate rotational symmetry,
as an 'intrinsic' state with $M=0$ would always lead to a state to be
observed in the lab frame with $K=0$ \cite{Ring:1980}.
One could also ask more physically: if $M$ is an internal quantum
number that has rather little to do with what can be observed in the
lab frame, why don't we use 
\be 
P^I_K = \sum_M f_M P^{I}_{KM}
\ee
as a projector with $f_M$ as additional variational parameters
(cf. e.g. \cite{Kerman:1977}) ? In
\cite{Kamlah:1968} it was argued that one should take all $f_M$ to be
equal to one, but whatever one does, one always ends up with the
problem that one has to evaluate matrix elements of operators acting upon
projectors that do not carry the quantum numbers that are projected
upon; to be concrete, one has e.g. to evaluate
\be
 \frac{\sum_{M,M'} \langle \phi | \vec{J} P^{I}_{M
     M'}  | \phi \rangle} {\sum_{M,M'} \langle \phi | P^{I}_{M M'} |  \phi
   \rangle }. 
\ee
One sees clearly that, unlike in the Abelian case, the projector matrix
elements do not usually cancel. A number of ways have been devised in
the literature to approximately evaluate these sums, but they are not
of interest here, since they cannot be transcribed to the field
theory case. We only want to point to the fact that the difficulties
in even the formulation of the projector do not come into play if
we want to project onto a state with $I=0$, since then there is only
one projector $P^0_{00}$ and no factors of $f$ can be chosen. This is
the case of the Yang-Mills projector since we have combined the
gluonic ($\Gamma$) and the matter ($\rho$) contribution into one set
of operators ($\mathcal{G}$) which now are supposed to form an
's-wave'. The problem of non-factorizability of projectors and other
operators will nevertheless also haunt us later on, too
(cf. sec. \ref{sec_diff_ext_charg}). \\
\subsection{General Remarks about Validity}
One last comment shall be made at this point on the validity of the
Kamlah expansion. It is a priori not clear that it is physically
sensible to perform the Kamlah expansion for a local symmetry. In nuclear
physics, the very deformed states usually do not come about through a very
deformed nucleus in space, but are due to a very large number of
participating nucleons \cite{Ring:1980}. In a gauge field theory, one
has usually the 
situation of a very small number of excitations at every point in
space. Therefore it is not clear whether it is sensible to use the
'large-deformation' expansion. However, it will turn out that the
Kamlah expansion performs very well in the cases we have considered.
\section{\label{sec_QED}Application to Quantized Electrodynamics}
In this section, we want to apply the general principles outlined above to
what we call electrodynamics - a quantized electromagnetic field interacting
with static sources. It is illuminating since we will see that in marked
contrast to nuclear physics, the first and second order expansions are
even qualitatively quite different. 
\subsection{Space of Allowed Wave Functionals}
In the remainder of this paper we will restrict ourselves to wave
functionals of Gaussian form
\be
\psi[\mathbf{A}] = \exp{\left\{-(\mathbf{A} - \bar{\mathbf{A}})_i
\left(\frac{1}{4} G^{-1}_{ij} - i \Sigma_{ij} \right)  (\mathbf{A} -
\bar{\mathbf{A}})_j+ i 
\bar{\mathbf{e}}_i (\mathbf{A} - \bar{\mathbf{A}})_i \right\} }
\label{eq_def_most_general_Gaussian_state} 
\ee
where we use a super-index notation: when we consider electrodynamics,
$i,j$ both include a spatial and 
a position index, which is summed or integrated over, respectively. In
the case of Yang-Mills theories the super-indices furthermore include
a colour index.
$G^{-1}, \Sigma, \bar{\mathbf{A}}, \bar{\mathbf{e}}$ are all purely real.
Further conventions and a short account of canonical quantization are
given in appendix \ref{app_conv}. 
\subsection{Simplified Projector}
In the case of electrodynamics the expression for the projector can be
simplified significantly compared to eq.(\ref{eq_def_gen_proj}). Since in
electrodynamics 
there is only one charge operator, which also commutes with the
Hamiltonian, the fermionic states can be chosen as eigenstates of the
charge operator, and since we want to prescribe a certain external
charge configuration we consider only states
(cf. eq.\,(\ref{def_coupled_states}) for the non-Abelian case) of the form
\be
| \text{coupled} \rangle = | \text{fermion}~ \tilde{\rho} \rangle |
  \text{photon} \rangle, \label{def_coupled_states_ED}
\ee 
with 
\be
\rho (\mathbf{x}) | \text{fermion}~ \tilde{\rho} \rangle = \tilde{\rho}
(\mathbf{x}) | \text{fermion}~ \tilde{\rho} \rangle. 
\ee
Since we only want to project operators that depend solely on the gauge
field part, we can use a simplified projector $\PP^{\tilde{\rho}}$
\bea
\PP^{\rho} | \text{coupled} \rangle &=& \int \mathcal{D} \phi\, e^{i \int
\phi(\Gamma - \rho)} | \text{fermion}~ \tilde{\rho} \rangle |
\text{photon} \rangle 
=  | \text{fermion}~ \tilde{\rho} \rangle \int \mathcal{D} \phi \,
e^{i \int \phi(\Gamma - \tilde{\rho})} | \text{photon} \rangle \nonumber \\
& = & | \text{fermion}~ \tilde{\rho} \rangle \PP^{\tilde{\rho}} |
\text{photon} \rangle  
\eea 
which does not depend on the fermion charge operator but directly on
the prescribed c-number external charge distribution. Furthermore,
the (pure gauge part of the) Gauss law operator is also simpler than
in Yang-Mills theories. It is given by
\be
\Gamma(\mathbf{x}) = \nabla_i \mathbf{\Pi}_i(\mathbf{x}).
\ee
\subsection{Kamlah Expansion to First Order}
We start from the expression (cf. eqs.\,(\ref{Abelian_Kamlah_exp1},
\ref{Abelian_Kamlah_exp2}))     
\be
\langle H e^{i \int \phi (\Gamma - \rho)} \rangle = A_0 \langle e^{i \int
  \phi (\Gamma - \rho)} \rangle  + A_1(\mathbf{y}) \langle
  \{\Gamma(\mathbf{y}) - \langle \Gamma(\mathbf{y}) \rangle \} \ e^{i \int
  \phi (\Gamma - \rho)} \rangle, \label{definingfirstorderphoton}
\ee
where we have again used  $\int \phi (\Gamma - \rho)$ as an abbreviation for
$\int d^3z\, \phi(\mathbf{z}) (\Gamma(\mathbf{z}) -   \rho(\mathbf{z}))$. A
useful notation that will be used throughout the remainder of this
section is 
\be
\Delta(\mathbf{x}) = \Gamma(\mathbf{x}) - \langle \Gamma(\mathbf{x}) \rangle
\hspace*{0.25em}, \hspace*{2.75em} \delta(\mathbf{x}) = \rho(\mathbf{x}) -
\langle \Gamma(\mathbf{x}) \rangle. 
\ee
Introducing the functional inverse $\Theta^{-1}(\mathbf{x},\mathbf{y})$ of
$\langle \Delta(\mathbf{x}) \Delta(\mathbf{y}) \rangle$, s.t.
\be
\Theta^{-1}(\mathbf{x}, \mathbf{z}) \langle \Delta(\mathbf{z})
\Delta(\mathbf{y}) \rangle = \delta_{\mathbf{x} \mathbf{y}} 
\ee
we obtain for the projected energy functional (details of the
procedure in general are outlined in app. \ref{app_deriv}, where it is
applied to the specific case of the second order Kamlah expansion)
\be
\left. \frac{\langle H \PP^{\rho} \rangle}{\langle \PP^{\rho} \rangle}
\first = \langle H \rangle + \langle H \Delta(\mathbf{x}) \rangle
\Theta^{-1}(\mathbf{x},\mathbf{y}) \delta(\mathbf{y}),
\ee
where we have introduced the notation $\left. A \firstt$ to
denote the fact that $A$ is evaluated using the Kamlah expansion to
first order.
Note that the derivation can be carried out for any gauge-invariant
  few-body operator $\mathcal{O}$, since there is nothing special about
  the Hamiltonian: 
\be
\left. 
\frac{\langle \mathcal{O} \PP^{\rho} \rangle}{\langle \PP^{\rho}
\rangle} \first = \langle \mathcal{O} \rangle + \langle \mathcal{O}
  \Delta(\mathbf{x}) \rangle \Theta^{-1}(\mathbf{x},\mathbf{y}) 
\delta(\mathbf{y}). \label{Kamlah_ED_firstorder_generalOp}
\ee 
This provides a first test of the validity of the first-order Kamlah
expansion. We may set $\mathcal{O} = \Delta(\mathbf{z})$, and obtain
\bea
&& \left. \frac{\langle \Delta(\mathbf{z}) \PP^{\rho} \rangle}{\langle
\PP^{\rho} \rangle} \first
\stackrel{eq.\,(\ref{Kamlah_ED_firstorder_generalOp})}{=}  
\langle \Delta(\mathbf{z}) \rangle + \langle \Delta(\mathbf{z}) 
  \Delta(\mathbf{x}) \rangle \,\Theta^{-1}(\mathbf{x},\mathbf{y})
  \delta(\mathbf{y}) = \rho(\mathbf{z}) - \langle \Gamma(\mathbf{z})
  \rangle, \, \nonumber \\ & \text{i.e.} & \, \left.
\frac{\langle \Gamma(\mathbf{z}) \PP^{\rho} \rangle}{\langle
\PP^{\rho} \rangle} \first
\stackrel{eq.\,(\ref{Kamlah_ED_firstorder_generalOp})}{=}  
\rho(\mathbf{z}), 
\eea
showing that the expression provided by the first order Kamlah expansion for
$\langle \Gamma(\mathbf{z}) \PP^{\rho} \rangle/\langle \PP^{\rho} \rangle$ 
is \textit{exact}. It doesn't stay that way, however, if one considers
e.g. $\mathcal{O} = \Delta({\mathbf{z}_1}) \Delta({\mathbf{z}_2})$ :
\be
\left. \frac{\langle  \Delta({\mathbf{z}_1}) \Delta({\mathbf{z}_2}) \PP^{\rho}
\rangle}{\langle \PP^{\rho} \rangle} \first
\stackrel{eq.\,(\ref{Kamlah_ED_firstorder_generalOp})}{=}  
\langle \Delta({\mathbf{z}_1}) \Delta({\mathbf{z}_2})\rangle + 
\langle \Delta({\mathbf{z}_1}) \Delta({\mathbf{z}_2}) 
  \Delta(\mathbf{x}) \rangle \Theta^{-1}(\mathbf{x},\mathbf{y})
  \delta(\mathbf{y}) = \langle \Delta({\mathbf{z}_1}) \Delta({\mathbf{z}_2})
  \rangle,
\label{Kamlah_firstorder_fail}
\ee
where we have used that $\langle \Delta({\mathbf{z}_1})
\Delta({\mathbf{z}_2}) \Delta(\mathbf{x}) \rangle = 0$, which is true
for every Gaussian state as given by
eq.\,(\ref{eq_def_most_general_Gaussian_state}).  We see that there are no 
corrections due to the projector for $\Delta({\mathbf{z}_1}) \Delta({\mathbf{
z}_2})$ by the first 
order Kamlah formula. This gives rise to the expectation that the 
energy will not be correctly projected, since it contains 
terms of the type $\langle \Gamma(\mathbf{x})  \Gamma(\mathbf{y}) \PP^{\rho}
\rangle$ (as we will see below, they arise  from the kinetic energy). \\
Let us now evaluate the projected energy. For this it is useful 
to note that under a variety of
assumptions\footnote{Either one may assume that the kernel in the Gaussian
  wave functional is purely real, i.e. $\Sigma = 0$, or that both $G^{-1}$ and
  $\Sigma$ are translation invariant, or that the  background field
$\bar{\mathbf{B}}_i = \epsilon_{ijk} \nabla_j \bar{\mathbf{A}}_k$ fulfils the
classical equations of motion.} 
\be
\langle \mathbf{B}_i(\mathbf{x}) \mathbf{B}_{j}(\mathbf{y}) \Gamma(\mathbf{z})
\rangle = \langle \mathbf{B}_i(\mathbf{x}) \mathbf{B}_j(\mathbf{y}) \rangle
\langle  \Gamma(\mathbf{z}) \rangle,  
\ee
i.e. $\langle \mathbf{B}_i(\mathbf{x}) \mathbf{B}_{j}(\mathbf{y}) \Delta(\mathbf{z}) \rangle =
0$. Physically it is 
plausible that the magnetic part of the energy should not be affected, since
it contains the transversal degrees of freedom only. For the electrical field
one obtains 
\be
\langle \mathbf{\Pi}_i(\mathbf{x}) \mathbf{\Pi}_i(\mathbf{x}) \Gamma(\mathbf{z})
\rangle = 2 \langle 
\mathbf{\Pi}_i(\mathbf{x}) \Delta(\mathbf{z}) \rangle 
\langle \mathbf{\Pi}_i(\mathbf{x}) \rangle + \langle \mathbf{\Pi}_i(\mathbf{x})
\mathbf{\Pi}_i(\mathbf{x}) 
\rangle \langle \Gamma(\mathbf{z}) 
\rangle,
\ee
which is valid even without integrating over $\mathbf{x}$. 
With this, 
the first order energy correction is given by
\be
\langle H \Delta(\mathbf{z}) \rangle = \langle \mathbf{\Pi}_i(\mathbf{x}) \Delta(\mathbf{z}) \rangle \langle \mathbf{\Pi}_i(\mathbf{x}) 
\rangle. 
\ee
One can make further progress by assuming that
both $G^{-1}$ and $\Sigma$ are translationally invariant; then one can write
\be
\langle \mathbf{\Pi}_i(\mathbf{x}) \mathbf{\Pi}_j(\mathbf{y}) \rangle_c = \frac{1}{4}
G^{-1}_{ij}(\mathbf{x},\mathbf{y}) + 4 (\Sigma G 
\Sigma)_{ij}(\mathbf{x},\mathbf{y}) = \int \frac{d^3p}{(2 \pi)^3} e^{i
\mathbf{p}.(\mathbf{x}-\mathbf{y})}
(\tilde{\sigma}_L^\mathbf{p} \PP^{L_\mathbf{p}}_{ij} +
\tilde{\sigma}_T^{\mathbf{p}} \PP^{T_\mathbf{p}}_{ij})  
\ee
which defines the longitudinal and transversal components
$\tilde{\sigma}_L^\mathbf{p}, \tilde{\sigma}_T^\mathbf{p}$, and where
we have used the abbreviation $\PP^{L_\mathbf{p}}_{ij}$ for $\mathbf{p}_i
\mathbf{p}_j/\mathbf{p}^2$; correspondingly $\PP^{T_\mathbf{p}}_{ij} =
\delta_{ij} - \PP^{L_\mathbf{p}}_{ij}$.  
This allows an explicit expression for $\Theta^{-1}(\mathbf{x},\mathbf{y})$: 
\be
\Theta^{-1}(\mathbf{x},\mathbf{y}) = \int \frac{d^3p}{(2 \pi)^3} e^{i
\mathbf{p}.(\mathbf{x}-\mathbf{y})}
\frac{1}{\mathbf{p}^2 \tilde{\sigma}_L^\mathbf{p}} 
\ee
and thus
\bea
\langle H \Delta(\mathbf{x_1}) \rangle
\Theta^{-1}(\mathbf{x}_1,\mathbf{y}) & = &  \langle
\mathbf{\Pi}_i(\mathbf{x}) \rangle \langle 
\mathbf{\Pi}_i(\mathbf{x}) \Delta({\mathbf{x}_1}) \rangle 
\Theta^{-1}(\mathbf{x}_1,\mathbf{y}) \nonumber \\ 
&=& \langle \Gamma(\mathbf{x}) \rangle  G_{\Delta}(\mathbf{x},
\mathbf{y}).  
\eea
$G_{\Delta}$ denotes the Green's function of the (ordinary) Laplace
operator, i.e. $-\nabla_i \nabla_i G_{\Delta}(\mathbf{x},
\mathbf{y}) = \delta_{\mathbf{x}\mathbf{y}}$. 
Then the projected energy reads
\be
\left. \frac{\langle H \PP^{\rho} \rangle}{\langle \PP^{\rho} \rangle} \first
= \langle H \rangle - \langle \Gamma(\mathbf{x}) \rangle
G_{\Delta}(\mathbf{x}, \mathbf{y})  \Big(\langle
\Gamma(\mathbf{y}) \rangle - \rho(\mathbf{y})\Big). 
\ee
We can make this expression even more transparent if we have a look
  at
$\langle \mathbf{\Pi}_{i}^L(\mathbf{x}) \rangle \langle \mathbf{\Pi}_{i}^L(\mathbf{x})
  \rangle$ where $\mathbf{\Pi}_{i}^L(\mathbf{x})$ denotes the longitudinal
  component of $\mathbf{\Pi}_i(\mathbf{x})$,
i.e. $\mathbf{\Pi}_{i}^L(\mathbf{x}) = \int d^3y\,
(\PP^L)_{ij}(\mathbf{x}, \mathbf{y}) \mathbf{\Pi}_j(\mathbf{y})$; for
$(\PP^L)_{ij}(\mathbf{x}, \mathbf{y})$
cf. eq.\,(\ref{App_ord_long_proj1}). One can derive the identity 
\be
\langle \mathbf{\Pi}_{i}^L(\mathbf{x}) \rangle \langle
  \mathbf{\Pi}_{i}^L(\mathbf{x}) \rangle  = \langle 
\Gamma(\mathbf{x}) \rangle G_{\Delta}(\mathbf{x}, \mathbf{y}) \langle
\Gamma(\mathbf{y}) \rangle, 
\ee
and therefore one can write 
\bea
\left. \frac{\langle H \PP^{\rho} \rangle}{\langle \PP^{\rho} \rangle}
\first &=&
\langle H \rangle^P - \frac{1}{2}  \left\{\langle \Gamma(\mathbf{x})
\rangle G_{\Delta}(\mathbf{x}, \mathbf{y}) \langle \Gamma(\mathbf{y})
\rangle - 2 \langle \Gamma(\mathbf{x}) \rangle G_{\Delta}(\mathbf{x},
\mathbf{y}) \rho(\mathbf{y})\right\} \\
&=& \langle H \rangle^P - \frac{1}{2} \left( \Gamma(\mathbf{x}) -
\rho(\mathbf{x}) \right) G_{\Delta}(\mathbf{x}, \mathbf{y})
\left( \Gamma(\mathbf{y}) - \rho(\mathbf{y}) \right) +
\frac{1}{2} \rho(\mathbf{x}) G_{\Delta}(\mathbf{x}, \mathbf{y})
\rho(\mathbf{y}).
\eea
$\langle H \rangle^P$ denotes here the part of the energy
expectation value which is independent of $\langle \mathbf{\Pi}^L \rangle$, but
which still
depends on $\langle \mathbf{\Pi}^L \mathbf{\Pi}^L \rangle_c$, where we
have introduced the abbreviation $\langle A B \rangle_c = \langle A B
\rangle - \langle A \rangle \langle B \rangle$. In formulas
this is expressed simply by $\langle H 
\rangle^P = \langle H \rangle - \frac{1}{2} \langle
\mathbf{\Pi}^L_i(\mathbf{x})  \rangle \langle
\mathbf{\Pi}^L_i(\mathbf{x})  \rangle$. 
Thus $\langle \Gamma(\mathbf{x}) \rangle$ is determined to be equal to
$\rho(\mathbf{x})$ at the stationary point, giving the correct charge
contribution to the energy 
\be
\frac{1}{2} \rho(\mathbf{x}) G_{\Delta}(\mathbf{x}, \mathbf{y})
\rho(\mathbf{y}).  
\ee
As can be seen by explicit computation, at the stationary point
$\langle \mathbf{\Pi}^L \mathbf{\Pi}^L \rangle_c$ is zero, s.t. the
energy depends only on the transversal degrees of freedom and on the charge
density. However, the outcome is quite in contrast to what 
we expected from the Kamlah expansion initially: since we have
computed the projected 
energy functional, (a) this should be completely independent of the gauge
dependent degrees of freedom; thus, it should not depend on $\langle
\mathbf{\Pi}^L \mathbf{\Pi}^L \rangle_c$, since in order to provide a
non-vanishing expectation value $\langle\mathbf{\Pi}^L \mathbf{\Pi}^L
\rangle_c$ the wave functional has to depend on $\mathbf{A}^L$;
and (b) the state was assumed to be
 strongly deformed, 
but, if we take the first-order
formalism seriously, at the stationary point our state is just \textit{not
  deformed}, since $\langle \mathbf{\Pi}^L \mathbf{\Pi}^L \rangle_c =
0$. These problems are overcome by performing the Kamlah expansion to
second order. Therefore,  we now turn to this.  
\subsection{Kamlah Expansion to Second Order} 
Details of the derivation of the energy functional are given in
app.\,\ref{app_elec_kam_2nd}. It has to be noted that the
factorization properties of Gaussian wave functionals,
cf. eq.\,(\ref{eq_def_most_general_Gaussian_state}), and
translational invariance of both $G^{-1}$ and $\Sigma$ are used
heavily in this derivation. We obtain for the projected energy
functional (using the notation $\Delta H = H - \langle H \rangle$): 
\bea
\left. \frac{\langle H \PP^{\rho} \rangle}{\langle \PP^{\rho} \rangle}
\second & = & \langle H
\rangle - \frac{1}{2} \langle \Delta H \Delta({\mathbf{x}_1})
\Delta({\mathbf{x}_2}) \rangle
\Theta^{-1}(\mathbf{x}_1,\mathbf{x}_2) \nonumber \\ && - \langle
\Delta H \Delta({\mathbf{x}_1}) \rangle
\Theta^{-1}(\mathbf{x}_1,\mathbf{y}_1)  
\langle \Gamma({\mathbf{y}_1}) - \rho({\mathbf{y}_1}) \rangle
\label{sec_order_en_func_ED}  \\ 
& & + \frac{1}{2} \langle \Delta H \Delta({\mathbf{x}_1})
\Delta({\mathbf{x}_2}) \rangle \Theta^{-1}(\mathbf{x}_1,\mathbf{y}_1)
\Theta^{-1}(\mathbf{x}_2,\mathbf{y}_2) \langle 
\Gamma({\mathbf{y}_1}) - \rho({\mathbf{y}_1}) \rangle \langle
\Gamma({\mathbf{y}_2}) -  \rho({\mathbf{y}_2}) \rangle, \nonumber 
\eea
where we have introduced the notation $\left. A \secondt$ to
denote the fact that $A$ is evaluated using the Kamlah expansion to
second order. As in the paragraph above, in the derivation of
eq.\,(\ref{sec_order_en_func_ED}) no special property of $H$ apart
from gauge invariance was used, and therefore the expression is valid
for every gauge-invariant few-body operator $\mathcal{O}$:
\bea
\left. \frac{\langle \mathcal{O} \PP^{\rho} \rangle}{\langle
\PP^{\rho} \rangle} \second & = & \langle \mathcal{O} 
\rangle - \frac{1}{2} \langle \Delta \mathcal{O}
\Delta({\mathbf{x}_1}) \Delta({\mathbf{x}_2})  \rangle
\Theta^{-1}(\mathbf{x}_1,\mathbf{x}_2)  \nonumber \\ & & 
\displaystyle - \langle \Delta \mathcal{O} \Delta({\mathbf{x}_1})
\rangle 
\Theta^{-1}(\mathbf{x}_1,\mathbf{y}_1) \langle \Gamma(\mathbf{y_1}) -
\rho({\mathbf{y}_1}) \rangle  \label{Kamlah_2_gen_op} \\ 
& & + \frac{1}{2} \langle \Delta \mathcal{O} \Delta(
{\mathbf{x}_1})
\Delta({\mathbf{x}_2}) \rangle \Theta^{-1}(\mathbf{x}_1,\mathbf{y}_1)
\Theta^{-1}(\mathbf{x}_2,\mathbf{y}_2) \langle 
\Gamma({\mathbf{y}_1}) - \rho({\mathbf{y}_1}) \rangle \langle \Gamma({\mathbf{y}_2}) -
\rho({\mathbf{y}_2}) \rangle. \nonumber 
\eea
As in the first order case, we now have a look at $\mathcal{O} =
\Delta(\mathbf{z})$ and 
$\mathcal{O} = \Delta({\mathbf{z}_1}) \Delta({\mathbf{z}_2})$:
\be
\left. \frac{\langle \Delta(\mathbf{z}) \PP^{\rho} \rangle}{\langle \PP^{\rho}
\rangle} \second \stackrel{eq.\,(\ref{Kamlah_2_gen_op})}{=}
\rho(\mathbf{z}) - \langle \Gamma(\mathbf{z}) \rangle   
\ee
which is the same result as in the first-order expansion, since all the
additional terms due to the second order vanish; next
\be
\left. \frac{\langle \Delta({\mathbf{z}_1}) \Delta({\mathbf{z}_2}) \PP^{\rho}
\rangle}{\langle \PP^{\rho} \rangle} \second
\stackrel{eq.\,(\ref{Kamlah_2_gen_op})}{=}  \langle
\Gamma(\mathbf{z}_1) - \rho({\mathbf{z}_1}) \rangle \langle 
\Gamma(\mathbf{z}_2) - \rho({\mathbf{z}_2})
\rangle. 
\ee
Thus, in the second-order Kamlah expansion, both $\langle \Delta(\mathbf{
z}) \PP^{\rho} \rangle/\langle \PP^{\rho} \rangle$ and $\langle
\Delta({\mathbf{z}_1}) \Delta({\mathbf{z}_2}) 
\PP^{\rho} \rangle/\langle \PP^{\rho} \rangle$ are
reproduced \textit{exactly} (cf. sec.\,\ref{kamlah_pattern_restore})
. This gives rise to the hope that the energy, too, will be 
projected correctly, since it depends both on $\langle \Gamma(\mathbf{x})
  \rangle \langle \Gamma(\mathbf{y}) \rangle$ and on  $\langle
  \Gamma(\mathbf{x})  \Gamma(\mathbf{y}) \rangle_c$.   
The detailed derivation of the simplified expression for the projected
  energy is given in app.\,\ref{app_elec_kam_2nd}. The main lessons
  are that, under the proposed assumptions, the magnetic energy does
  not obtain a correction from the projector (since it depends on
  transversal degrees of freedom only) whereas the correction of the
  Kamlah expansion subtracts just the longitudinal part of the
  electric energy; one obtains eventually a  concise formula for the
  projected energy: 
\be
\left. \frac{\langle H \PP^{\rho} \rangle}{\langle \PP^{\rho} \rangle} \second
=  \langle H  \rangle - \frac{1}{2} \langle \PP^L_{ji}(\mathbf{y}, \mathbf{x})
\mathbf{\Pi}_i (\mathbf{x}) \mathbf{\Pi}_j (\mathbf{y})
\rangle + \frac{1}{2} \rho(\mathbf{x}) G_{\Delta} (\mathbf{x},
\mathbf{y}) \rho(\mathbf{y}). 
\label{ED_proj_energ}  
\ee
Note how different the outcome is from the first-order calculation: the
projected energy is completely independent of the longitudinal parts of the
electric field (both the expectation value and the correlator), and in turn
they cannot be determined variationally. In the wave functional, the
longitudinal parts are totally
undetermined, which is what one wants since their value is prescribed by the
projector. Thus we can conclude that, at least in electrodynamics, the second
order Kamlah expansion is mandatory for a consistent formalism.  \\
One amusing point might be noted at the end of this section; if one
restricts the dimensionality of space-time to (1+1), there are no
transversal components of $\mathbf{A}$, $\mathbf{\Pi}$; thus the projection
to second order as carried out by the Kamlah expansion leaves behind
only the interaction of static charges. Though we have not
checked explicitly (by replacing Fourier integrals by sums etc),
eq.\,(\ref{ED_proj_energ}) seems to suggest strongly that if we
consider the theory on a compact spatial manifold in (1+1) dimensions,
we would also keep - in addition to the energy of the charges - 
the spatially constant part of $\mathbf{\Pi}$
(since this part is annihilated by the derivatives in the longitudinal
projector, and is thus not subtracted), just as one would expect.
\subsection{General Observations \label{kamlah_pattern_restore}}
A certain pattern
  could be observed when going from the first- to the second-order
  expansion: in the first-order expansion the expectation 
  value of a single Gauss law operator
  between projected states was corrected, but not those of more
  operators (we checked the expectation value of $\Delta(\mathbf{z_1})
  \Delta(\mathbf{z}_2)$ in eq.\,(\ref{Kamlah_firstorder_fail})). In the
  second-order expansion the expectation values of up to two Gauss law
  operators
  between projected states were corrected, but not those of more
  operators (it is easy to convince oneself that $\langle
  \Delta(\mathbf{z}_1) \Delta(\mathbf{z}_2) \Delta(\mathbf{z}_3)
  \PP^{\rho}\rangle / \langle \PP^{\rho} \rangle
  \stackrel{eq.\,(\ref{Kamlah_2_gen_op})}{=} - \langle 
  \Delta(\mathbf{z}_1) \Delta(\mathbf{z}_2) \rangle \langle
  \Gamma(\mathbf{z}_3) - \rho(\mathbf{z}_3) \rangle +
  \text{permutations}$). Thus, we would 
  expect that in order to have the expectation values of up to three Gauss law
  operators corrected we need a third-order expansion etc, although we
  have not checked this explicitly. To what extent this pattern
  extends also to other operators is currently not known. The Gauss
  law operator is in this context obviously useful as a first check on
  the accuracy of the method, since one knows what to expect of the
  projected expectation value beforehand. One can give a further argument in
  support of the pattern found here: If we want to project correctly
  the product of $n$ Gauss law operators we have to obtain a product
  of $n$ charge distributions $\rho$ from the Kamlah expression for
  the projected operator. We have seen that the
  coefficients $A_i$ are independent of $\rho$ so the only place where
  $\rho$ does appear is in the terms $(\langle \Gamma \rangle -
  \rho)^i$ (where for simplicity we haven't written out the
  product). But the highest power $i$ that we obtain is equal to the
  order of the Kamlah expansion, i.e. a Kamlah expansion to second
  order includes at most terms $\propto \rho^2$. Thus the operator
  $\Gamma^3$ can never be correctly projected to give $\rho^3$ in the
  second order Kamlah expansion and a third order treatment would be
  called for.
\section{\label{sec_YMT}Application to Abelianized Yang-Mills Theories}
\subsection{Difficulties with External Charges \label{sec_diff_ext_charg}}
After the experience with the Kamlah expansion in electrodynamics,
we want to try to apply it also to the Yang-Mills case. We will see,
however, that we cannot incorporate charges as simply as in 
electrodynamics, and an explicit computation does not transcend the
leading order in perturbation theory (although a possible background field is
treated to all orders). \\
In order to discuss
why we cannot include charges, the first order of the Kamlah expansion
is sufficient; we don't consider the first order any further than
necessary for this purpose, 
since its shortcomings were already obvious in the 
discussion of the electrodynamic case. Again we start from the
expression (cf. eqs.\,(\ref{Abelian_Kamlah_exp1}, \ref{Abelian_Kamlah_exp2}))
\be
\langle H e^{-i \int \varphi^a \mathcal{G}^a} \rangle = 
A_0 \langle  e^{-i \int \varphi^a \mathcal{G}^a} \rangle  + 
A_1^a(\mathbf{y})  
\langle \Delta^a(\mathbf{y}) e^{-i \int \varphi^a \mathcal{G}^a} \rangle
\label{YMKamlah1stansatz} 
\ee
where we have used the abbreviation $\Delta^a(\mathbf{x}) = \Gamma^a(\mathbf{x}) -
\langle \Gamma^a(\mathbf{x}) \rangle$; later on, we will also
use\footnote{Here, a little care has to be taken w.r.t. the placement
of $g$s. If we use 'perturbative scaling', cf. app. \ref{app_conv},
the generator of gauge transformations in the 
gluonic sector is not $-\Gamma^a$, but $-\frac{1}{g} \Gamma^a$. Thus the
projector is in principle $\int \mathcal{D} \varphi\, e^{-i \int g \varphi^a
\mathcal{G}^a} = \int \mathcal{D} \varphi\, e^{i \int \varphi^a (\Gamma^a - g
\rho^a)}$. However, this explicit scaling will only be used later on
when we have already done away with the charges; we use the 'non-perturbative
scaling' while we are developing most of the formalism, and
re-install explicit factors of $g$ later on.} $\delta^a(\mathbf{x})
= \rho^a(\mathbf{x}) - \langle \Gamma^a(\mathbf{x}) \rangle$. In contrast to the
previous section, $\delta^a(\mathbf{x})$ is now an \textit{operator}. 
Using the procedure that is outlined in app. \ref{app_elec_kam_2nd},
we obtain
\be
\left. \langle H \PP^{\rho} \rangle \first = \langle H \rangle \langle
\PP^{\rho} \rangle +   \langle \text{YM} | H \Gamma^a(\mathbf{x}) |\text{YM}
\rangle (\Theta^{-1})^{ab}(\mathbf{x},\mathbf{y})  
\langle \delta^b(\mathbf{y}) \PP^{\rho} \rangle,
\ee
where we have used that $\Gamma^a(\mathbf{x}) \PP^{\rho} = \rho^a(\mathbf{x})
\PP^{\rho}$, and that our states $| \rangle$ have the form $|
\text{fermion} \rangle | \text{YM} \rangle$ (This is actually a
restriction - cf. eq.\,(\ref{def_coupled_states}) - which is of no 
importance to the argument made here). At this point we see already that the
results of electrodynamics 
do not generalize straightforwardly: there the fermionic states could
be chosen as eigenstates of \textit{all} charge operators $\rho$. In
Yang-Mills theories this is 
not possible (apart from states which are either non-charged or where
the charges form locally colour-singlets),
and therefore we \textit{cannot} write
\be
\langle \delta^b(\mathbf{y}) \PP^{\rho} \rangle = \langle
\delta^b(\mathbf{y}) \rangle \langle \PP^{\rho} \rangle. 
\ee
Thus, we cannot get rid of all expectation values containing the
projector simply by considering the normalized projected energy
expectation value 
$\langle H \PP^{\rho} \rangle/\langle \PP^{\rho} \rangle$. This,
however, was the main appeal of the Kamlah expansion, and in order to
rescue as much as possible of it, we will consider only uncharged
states in the following: states that satisfy $\langle f| \rho^a(\mathbf{x}) =
0$ and $\rho^a(\mathbf{x}) | f \rangle = 0$. In this case, the
projector is reduced to its purely gluonic part, and therefore called
$\PP$ in the remainder of this section. 
\subsection{Kamlah Expansion to Second Order \label{sec_YM_KE_2nd_order}}
We will now turn to the second order Kamlah expansion. While 
determining the coefficients $A_0, A_1, A_2$, we still work with general
(possibly charged) states; it will turn out that the coefficients are
actually independent of whether we have a charged state or not. We
have set $\Sigma \equiv 0$ throughout the whole
calculation. Then, we start from
(cf. eqs.\,(\ref{Abelian_Kamlah_exp1},\ref{Abelian_Kamlah_exp2}))  
\be
\langle H e^{-i \int \varphi^a \mathcal{G}^a} \rangle = 
A_0 \langle  e^{-i \int \varphi^a \mathcal{G}^a} \rangle  + 
A_1^a(\mathbf{y}) \langle \Delta^a(\mathbf{y}) e^{-i \int \varphi^a {\cal
G}^a} \rangle +  A_2^{ab}(\mathbf{y},\mathbf{z}) 
\langle \Delta^a(\mathbf{y}) \Delta^b(\mathbf{z})  e^{-i \int \varphi^a
\mathcal{G}^a} \rangle. 
\label{YMKamlah2ndansatz} 
\ee
It seems to be quite sensible to require that
$A_2^{ab}(\mathbf{y},\mathbf{z})$ should be   
symmetric under interchange of all indices, i.e. $A_2^{ab}(\mathbf{y},\mathbf{z}) = 
A_2^{ba}(\mathbf{z},\mathbf{y})$, since (a) most of $\Delta^a(\mathbf{y})
\Delta^b(\mathbf{z})$ is 
symmetric apart from $\Gamma^a(\mathbf{y}) \Gamma^b(\mathbf{z})$
\footnote{Note that $[\Gamma^a, \Gamma^b] \neq 0$,
cf. eq.\,(\ref{eq_comm_rel_Gauss_law}).}, (b) the antisymmetric 
part of $\Delta^a(\mathbf{y}) \Delta^b(\mathbf{z})$ is  $-\frac{i}{2} f^{abc}
\Gamma^c(\mathbf{y}) 
\delta_{\mathbf{y} \mathbf{z}}$ and should therefore rather be assigned to the term
multiplying $A_1$ from a systematic point of view. 
In the course of the calculation, we have to perform two approximations:
first, we have to restrict ourselves to Gaussian states
(as in electrodynamics),
cf. eq.\,(\ref{eq_def_most_general_Gaussian_state}). However, we have
to make a second 
approximation, which is more severe: we will restrict the (gluonic
part of the) Gauss law operator to its $\mathcal{O}(g^0)$ part.
In order to facilitate this perturbative treatment, 
we start with the 'non-perturbative
scaling' scheme (cf. table \ref{tab_pert_nonpert_scaling} in
app. \ref{App_scaling}) and then replace $\mathbf{A}$ by
$\bar{\mathbf{A}} + g \mathbf{a}$. Then, $\mathbf{a}$ is the new
dynamical quantum field 
with $\langle \mathbf{a} \rangle = 0$, and $\frac{1}{i} \frac{\delta}
{\delta \mathbf{a}}$ is in the following denoted by $\mathbf{\Pi}$.
The pair $(\mathbf{a}, \mathbf{\Pi})$ then follows the 'perturbative
scaling' scheme. Note that in this scaling scheme we continue to call
$\hat{\mathbf{D}}{}^{ab}_i(\mathbf{x}) \mathbf{\Pi}^b_i(\mathbf{x})$
the (gluonic part of the) Gauss law operator, even though the
generator of gauge transformations in this scheme is $-\frac{1}{g}
\hat{\mathbf{D}}{}^{ab}_i(\mathbf{x}) \mathbf{\Pi}^b_i(\mathbf{x})$.
The Gauss law operator is now expanded w.r.t. $g$ as follows:
\bea
\Gamma^a(\mathbf{x}) &=& \hat{\mathbf{D}}{}^{ab}_i(\mathbf{x})
\mathbf{\Pi}^b_i(\mathbf{x}) = 
(\hat{\bar{\mathbf{D}}}{}^{ab}_i(\mathbf{x})  - g f^{acb} \mathbf{
a}^c_i(\mathbf{x}) ) \mathbf{\Pi}_i^b(\mathbf{x}) = \bar{\Gamma}^a(\mathbf{x}) - g
f^{acb} \mathbf{a}^c_i(\mathbf{x}) 
\mathbf{\Pi}^b_i(\mathbf{x}) \nonumber \\ 
&=& \bar{\Gamma}^a(\mathbf{x}) + \mathcal{O}(g),
\eea
i.e. $\bar{\Gamma}^a(\mathbf{x}) = \hat{\bar{\mathbf{D}}}{}^{ab}_i(\mathbf{x})
\mathbf{\Pi}^b_i(\mathbf{x})$, where
$\hat{\bar{\mathbf{D}}}{}^{ab}_i(\mathbf{x})$ is the covariant
derivative in the background field $\bar{\mathbf{A}}$, explicitly
$\hat{\bar{\mathbf{D}}}{}^{ab}_i(\mathbf{x}) = \nabla_i \delta^{ab} -
f^{acb} \bar{\mathbf{A}}_i(\mathbf{x})$ .  
This then will allow to use factorizability in a fashion identical to
the case of electrodynamics. \\
However, this approximation has consequences; the first one is that we
have reduced the non-Abelian SU(N) symmetry,
cf. eq.\,(\ref{eq_comm_rel_Gauss_law}), to an Abelian $U(1)^{(N^2-1)}$ 
symmetry since: 
\be
~[\bar{\Gamma}^a(\mathbf{x}), \bar{\Gamma}^b(\mathbf{y}) ] = 0.
\ee
The second one has to do with the admissible background fields
$\bar{\mathbf{A}}$; it is clear that we cannot expect accuracy of the energy
expectation value to orders higher than $\mathcal{O}(g^0)$; thus we 
approximate the Hamiltonian by its $\mathcal{O}(g^0)$
contribution\footnote{This approximation is also sensible from another
point of view: the $\mathcal{O}(g^0)$ contribution of the Hamiltonian is 
at most quadratic in the operators $\mathbf{A}$ and $\mathbf{\Pi}$ and from the
case of electrodynamics one expects that the second-order Kamlah
expansion treats correctly only (projected) expectation values of up
to two operators, cf. sec. \ref{kamlah_pattern_restore}.}. The
kinetic energy remains unchanged, but the magnetic 
energy is simplified quite dramatically if we insert the decomposition
$\mathbf{A}^a_i(\mathbf{x}) = \bar{\mathbf{A}}^a_i(\mathbf{x}) + g
\mathbf{a}^a_i(\mathbf{x})$ into  
$\frac{1}{g^2}  \mathbf{B}^a_i(\mathbf{x}) 
\mathbf{B}^a_i(\mathbf{x})$ and neglect all terms of higher order than
$g^0$ :
\bea & & 
\frac{1}{g^2} \mathbf{B}^a_i(\mathbf{x}) \mathbf{B}^a_i(\mathbf{x}) \nonumber \\
&=& \frac{1}{g^2} 
\bar{\mathbf{B}}^a_i(\mathbf{x}) 
\bar{\mathbf{B}}^a_i(\mathbf{x}) + \frac{2}{g} \bar{\mathbf{B}}^a_i(\mathbf{x})
\epsilon_{i j_1 k_1} 
(\hat{\bar{\mathbf{D}}}{}^{a c_1}_{j_1}(\mathbf{x}) \delta_{\mathbf{x} \mathbf{z}_1})
 \mathbf{a}^{c_1}_{k_1}(\mathbf{z}_1) \label{MFexpansion0}
 \\ & & +
\left( \epsilon_{i j_1 k_1}
(\hat{\bar{\mathbf{D}}}{}^{a c_1}_{j_1}(\mathbf{x}) \delta_{\mathbf{x} \mathbf{z}_1})
 \epsilon_{i j_2 k_2}
(\hat{\bar{\mathbf{D}}}{}^{a c_2}_{j_2}(\mathbf{x}) \delta_{\mathbf{x} \mathbf{z}_2})
- \bar{\mathbf{B}}^a_i(\mathbf{x}) f^{a c_1 c_2} \epsilon_{i k_1 k_2}
 \delta_{\mathbf{x} \mathbf{z}_1} 
 \delta_{\mathbf{x} \mathbf{z}_2} \right) \mathbf{a}^{c_1}_{k_1}(\mathbf{z}_1)
 \mathbf{a}^{c_2}_{k_2}(\mathbf{z}_2). \nonumber \\
& = & \frac{1}{g^2} \bar{\mathbf{B}}^a_i(\mathbf{x}) \bar{\mathbf{B}}^a_i(\mathbf{x}) + \frac{2}{g} 
 J^{c_1}_{k_1}(\mathbf{x};\mathbf{z}_1) \mathbf{a}^{c_1}_{k_1}(\mathbf{z}_1)
+ M^{c_1 c_2}_{k_1 k_2} (\mathbf{x};\mathbf{z}_1,\mathbf{z}_2)
 \mathbf{a}^{c_1}_{k_1}(\mathbf{z}_1) \mathbf{a}^{c_2}_{k_2}(\mathbf{z}_2). 
\label{MFexpansion}
\eea
For eqs.\,(\ref{MFexpansion0}, \ref{MFexpansion}) to be valid we do not
have to integrate over $\mathbf{x}$. 
The restriction on $\bar{\mathbf{A}}$ comes about from the requirement that the
\textit{approximated Hamiltonian} should commute with the \textit{approximated Gauss
law operator}, 
otherwise the (implicit) assumption of the  Kamlah expansion - that
in the beginning the Hamiltonian commutes with the projector - is
incorrect. That one obtains a non-trivial condition at all results
from the fact that, to $\mathcal{O}(g^0)$ of the commutator of the full
quantities, one gets also a cross term from the $\mathcal{O}(g^{-1})$ part of
$\mathbf{B}^2$ with the $\mathcal{O}(g^1)$ part of $\Gamma$. Thus we
can express the requirement of gauge invariance of the approximated
Hamiltonian in two different ways: 
\begin{itemize} \item[(a)]  the term of order $g^{-1}$ in
eq.\,(\ref{MFexpansion}) should vanish, since then there can be no cross
term with the $\mathcal{O}(g^1)$ contribution to $\Gamma$. With a partial
integration, we see that this condition can be written as
\be
 \epsilon_{i j_1 k_1}
\hat{\bar{\mathbf{D}}}{}^{c_1 a}_{j_1}(\mathbf{x}) \bar{\mathbf{B}}^a_i(\mathbf{x}) =
- \hat{\bar{\mathbf{D}}}{}^{c_1  a}_{j_1}(\mathbf{x}) F_{j_1
  k_1}^a(\mathbf{x}) \stackrel{!}{=} 0,  
\ee
which is nothing but the classical equation of motion for the (static)
Yang-Mills field, since $F$ is the field strength tensor.
\item[(b)] the commutator $\frac{1}{g^2} [\mathbf{B}^2, \bar{\Gamma}]$
should vanish. We can compute the commutator and obtain
\be
\frac{1}{g^2} [\mathbf{B}^a_i(\mathbf{x}) \mathbf{B}^a_i(\mathbf{x}),
\bar{\Gamma}^b(\mathbf{y})] = 2 
\hat{\bar{\mathbf{D}}}{}^{b b_1}_{l_1}(\mathbf{y}) M^{b_1 c_2}_{l_1 k_2} (\mathbf{x};\mathbf{y},\mathbf{z}_2) 
\mathbf{a}^{c_2}_{k_2}(\mathbf{z}_2) \stackrel{!}{=} 0,
\ee
where we have used the fact that $M$ - as defined by
eqs.\,(\ref{MFexpansion0}, \ref{MFexpansion}) - is symmetric under
interchange of 
all its indices. This is of course equivalent to condition (a),
and thus to the requirement of the background field fulfilling the
classical equations of motion for static fields\footnote{In the
electrodynamical case there 
is no $\mathcal{O}(g^1)$ contribution to the Gauss law operator and hence
the whole discussion is not necessary.}.
\end{itemize}
The computation is given in some detail in app. \ref{app_YM_kam_2nd_detail}.  
Here we want to mention that in the course of this computation a
further restriction has to be made of the class of admissible kernels
$G^{-1}$ which is stated explicitly in eq.\,(\ref{restr_on_G_YM}). We
then obtain an overall correction to the electric energy  
\be
- \frac{1}{2} \int d^3x_1\, d^3x_2\, (\Pi_L)^{c_2 c_1}_{i_2i_1}
(\mathbf{x}_2,\mathbf{x}_1) \langle
\mathbf{\Pi}^{c_1}_{i_1}(\mathbf{x}_1) \mathbf{\Pi}^{c_2}_{i_2}(\mathbf{x}_2)
\label{Kamlah_overall_corr_electr_ener} 
\rangle,
\ee
where $\Pi_L$ is the (generalized) longitudinal projector as defined
in app. \ref{App_proj}.
Thus, the electrical energy neither contains
$\hat{\bar{\mathbf{D}}}{}^{ab}_i \bar{\mathbf{e}}^b_i$ nor the
(generalized) longitudinal component of $G^{-1}$. Since by requirement
of approximate gauge invariance the magnetic energy is transversal as
well, this is true for the total energy, and to $\mathcal{O}(g^0)$, we
can write  
\bea 
\left.
\frac{\langle H \PP \rangle}{\langle \PP \rangle}\second \quad =&&
\frac{1}{2} \int d^3x_1\, d^3x_2\,  (\Pi_T)^{c_2 c_1}_{i_2i_1} 
(\mathbf{x}_2,\mathbf{x}_1) \langle
\mathbf{\Pi}^{c_1}_{i_1}(\mathbf{x}_1)
\mathbf{\Pi}^{c_2}_{i_2}(\mathbf{x}_2)  
\rangle \nonumber \\ & +& \frac{1}{2} \int d^3x 
\left\{\frac{1}{g^2} \bar{\mathbf{B}}^a_i(\mathbf{x}) \bar{\mathbf{B}}^a_i(\mathbf{x}) + \tr{(\mathcal{M} G_{\mathbf{x} \mathbf{y}})|_{\mathbf{y} \rightarrow \mathbf{x}}} \right\}, 
\eea
with 
$\tr{(\mathcal{M} G_{\mathbf{x} \mathbf{y}})} =
(({\check{\bar{\mathbf{D}}}}(\mathbf{x}){\check{\bar{\mathbf{D}}}}(\mathbf{x}) 
- \check{\bar{\mathbf{B}}}(\mathbf{x}))^{ab}_{ij}
G^{ba}_{ji}({\mathbf{x}, \mathbf{y}}))$, 
where we have used the abbreviations
$\check{\bar{\mathbf{D}}}^{ab}_{ij}(\mathbf{x}) = \epsilon_{ikj}
{\hat{\bar{\mathbf{D}}}}{}^{ab}_k(\mathbf{x}),
\check{\bar{\mathbf{B}}}^{ab}_{ij}(\mathbf{x}) =
{\bar{\mathbf{B}}}^c_k(\mathbf{x}) f^{acb} \epsilon_{ikj}$;
for $\Pi_T$ cf. eq.\,(\ref{eq_def_Pi_T}). 
Writing the kinetic energy also in terms of $G^{-1}$ and
$\bar{\mathbf{e}}$, the energy is given as 
\bea
E^{proj}_{approx} &=& \left. \frac{\langle H \PP \rangle}{\langle \PP
\rangle} \second = \quad \frac{1}{2} \int 
d^3x_1\, d^3x_2\, \left\{ (\Pi_T)^{c_2 c_1}_{i_2i_1}
(\mathbf{x}_2,\mathbf{x}_1) \bar{\mathbf{e}}^{c_1}_{i_1}(\mathbf{x}_1)
\bar{\mathbf{e}}^{c_2}_{i_2}(\mathbf{x}_2) \right\}  + \frac{1}{8}
\Tr{(G^{-1}_{TT})} \nonumber \\ & & \qquad \qquad \quad + 
 \frac{1}{2} \int d^3x 
 \left\{ \frac{1}{g^2}\bar{\mathbf{B}}^a_i(\mathbf{x}) \bar{\mathbf{B}}^a_i(\mathbf{x}) +
  \tr{(\mathcal{M} G_{\mathbf{x} \mathbf{y}})|_{\mathbf{y} \rightarrow
 \mathbf{x}}} \right\}, \label{en_kamlah_2nd_expl} 
\eea
where $G^{-1}_{TT}$ is defined by eq.\,(\ref{restr_on_G_YM}), and the
trace $\Tr$ furthermore runs over colour, spatial and position indices.
We see that the projected energy is in fact independent of the
(generalized) longitudinal components of $\bar{\mathbf{e}}$ and $G^{-1}$. This
is what we would expect from the projector: we treat it to
$\mathcal{O}(g^0)$, and we obtain an energy functional that is compatible
with the energy functional of a state that is annihilated to ${\cal
O}(g^0)$ by the Gauss law operator, cf. app. \ref{app_PT}. Perturbation 
theory thus seems not to invalidate the treatment of the Kamlah
expansion. It is interesting to note that
eq.\,(\ref{en_kamlah_2nd_expl}) is  (at least formally) 
identical to the improved energy functional obtained in 
\cite{Heinemann:1999ja} in the case of magnetic background fields,
to the same order in perturbation theory as considered here. One
difference, however, should be noted: in our treatment the 
condition that $\bar{\mathbf{A}}$ has to satisfy the classical equations of
motion (and thus $\mathcal{M}$ has to be transversal in the generalized sense)
occurred naturally due to 
the requirement that the approximated Hamiltonian should still commute
with the approximated Gauss law operator, whereas in
\cite{Heinemann:1999ja}
it is apparently assumed that $\mathcal{M}$ is transversal in the
generalized sense. In the application considered in
\cite{Heinemann:1999ja}, the Savvidy vacuum, the condition is of
course fulfilled. In other cases, as e.g. the constant magnetic field
stemming from a non-commuting gauge potential \cite{Brown:1979bv,
Huang:1994vx}, this need not be the case. \\
As a last topic in this section we want to consider the value of the
energy if we consider it at the stationary point of $\bar{\mathbf{e}}$ and $G$.
We first consider $\bar{\mathbf{e}}$:
\be
\frac{\delta}{\delta \bar{\mathbf{e}}^a_i(\mathbf{x})} E^{proj}_{approx} = \int
d^3x_1\, \left\{ (\Pi_T)^{a c_1}_{i i_1} (\mathbf{x},\mathbf{x}_1)
\bar{\mathbf{e}}^{c_1}_{i_1}(\mathbf{x}_1) \right\} \stackrel{!}{=} 0,
\ee
thus the transversal component of $\bar{\mathbf{e}}$ has to be zero at the
stationary point. Next we consider $G_{TT}$ (since $\mathcal{M}$ is transversal,
the magnetic energy also contains only $G_{TT}$). For the variation
w.r.t. $G_{TT}$ we need the relation
\be
\frac{\delta}{\delta (G_{TT})^{a_1 a_2}_{i_1 i_2}(\mathbf{z}_1,\mathbf{z}_2)}
\Tr{(G^{-1})} 
= - (G^{-1}_{TT} G^{-1}_{TT})^{a_1 a_2}_{i_1 i_2}(\mathbf{z}_1,\mathbf{z}_2). 
\ee
After a short calculation we end up with 
\be
E^{proj}_{approx}[\bar{\mathbf{A}}] = 
\frac{1}{2}
\Tr{\Big((\check{\bar{\mathbf{D}}} \check{\bar{\mathbf{D}}} -
\check{\bar{\mathbf{B}}})^{\frac{1}{2}}\Big)}  + \frac{1}{2 g^2} \int d^3x\,
\bar{\mathbf{B}}^a_i(\mathbf{x}) \bar{\mathbf{B}}^a_i(\mathbf{x}).
\label{Kamlah_YM_final_energy1}  
\ee
Thus we see that at the stationary point, the difference to the 
treatment of \cite{Kerman:1989kb}, where the projector was ignored,
lies only in the 
requirement that here $\bar{\mathbf{A}}$ has to satisfy the classical
equations of motion. The important point of the Kamlah expansion is,
however, that the energy functional depends only on the transversal degrees
of freedom even away from the stationary point.
\subsection{Treatment of Charges to Lowest Order in $g$}
Here we only want to insert an additional point that is inspired by
the observation that, in the leading-order perturbative treatment, the gauge
group $SU(N)$ is reduced to a direct product of $U(1)$ groups. In that
case one can obviously also require the charged states to be
simultaneous eigenstates of all charge operators, and can therefore
factorize 
\be
\langle \delta^b(\mathbf{y}) \PP^{\rho} \rangle = \delta^b(\mathbf{y})
\langle \PP^{\rho} \rangle, 
\ee
where the $\delta^b(\mathbf{y})$ on the RHS is no longer an operator but simply
a c-number function. At this point, we should make clear that in the
'perturbative scaling', cf. app. \ref{app_conv}, that has been used
here, $\delta$ reads 
\be
\delta^a(\mathbf{x}) =  g \rho^a(\mathbf{x}) - \langle
\bar{\Gamma}^a(\mathbf{x}) \rangle, 
\ee
since Gauss' law implies $\mathcal{G}^a | \rangle = (-\frac{1}{g}
\Gamma^a + \rho^a) | \rangle = 0 \rightarrow \Gamma^a | \rangle =  g
\rho^a | \rangle$. For this kind of quasi-Abelian charges, we can
immediately obtain the projected energy to second order in the Kamlah
expansion by replacing $\langle \bar{\Gamma}^a \rangle$ by $\langle
\bar{\Gamma}^a \rangle - g \rho^a$ in eq.\,(\ref{YM_Kamlah_final1}):
\bea
\left.
\frac{\langle H \PP^{\rho} \rangle}{\langle \PP^{\rho} \rangle} \second
&=& \langle H \rangle - \frac{1}{2}  \langle \Delta H
\bar{\Delta}^{c_1}(\mathbf{x}_1) 
\bar{\Delta}^{c_2}(\mathbf{x}_2) \rangle \Theta^{c_2
c_1}(\mathbf{x}_2,\mathbf{x}_1) \nonumber \\  
& & \hspace*{1.875em} -  \langle \Delta H \bar{\Delta}^{c_1}(\mathbf{x}_1) 
 \rangle \Theta^{c_2 c_1}(\mathbf{x}_2,\mathbf{x}_1) \Big(\langle
\bar{\Gamma}^{c_2}(\mathbf{x}_2) \rangle - g \rho^{c_2}(\mathbf{x}_2) \Big)
\label{YM_Kamlah_final2} \\ & &  \hspace*{1.875em} + 
\frac{1}{2} \langle \Delta H
 \bar{\Delta}^{c_1}(\mathbf{x}_1) 
\bar{\Delta}^{c_2}(\mathbf{x}_2) \rangle \Theta^{c_1
 d_1}(\mathbf{x}_1,\mathbf{y}_1)  
\Theta^{c_2 d_2}(\mathbf{x}_2,\mathbf{y}_2) \times \nonumber \\ & &
\hspace*{12em}\Big (\langle
\bar{\Gamma}^{d_1}(\mathbf{y}_1) \rangle - g \rho^{d_1}(\mathbf{y}_1) \Big)
\Big(\langle 
\bar{\Gamma}^{d_2}(\mathbf{y}_2)\rangle - g \rho^{d_2}(\mathbf{y}_2) \Big)
. \nonumber 
\eea
We see that it differs from the energy projected onto the chargeless
sector by
\bea
&& \hspace*{-0.75em} 
-g \Big\{ \left(
\langle \Delta H  \bar{\Delta}^{c_1}(\mathbf{x}_1)
\bar{\Delta}^{c_2}(\mathbf{x}_2) \rangle \Theta^{c_1
d_1}(\mathbf{x}_1,\mathbf{y}_1) \Theta^{c_2
d_2}(\mathbf{x}_2,\mathbf{y}_2)  \langle \bar{\Gamma}^{d_1}(\mathbf{y}_1) 
\rangle \right) \nonumber \\ & & \hspace*{8em} 
- \langle \Delta H \bar{\Delta}^{c_1}(\mathbf{x}_1)  
 \rangle \Theta^{c_1 d_2}(\mathbf{x}_1,\mathbf{y}_2) \Big\}
\rho^{d_2}(\mathbf{y}_2)  
 \\ &+&
\frac{g^2}{2} \langle \Delta H
 \bar{\Delta}^{c_1}(\mathbf{x}_1) \bar{\Delta}^{c_2}(\mathbf{x}_2) \rangle
 \Theta^{c_1 
 d_1}(\mathbf{x}_1,\mathbf{y}_1)  \Theta^{c_2 d_2}(\mathbf{x}_2,\mathbf{y}_2)
 \rho^{d_1}(\mathbf{y}_1)  \rho^{d_2}(\mathbf{y}_2).  \nonumber
\eea
The first term can be easily calculated to give  \textit{zero}
\footnote{We have already indicated above that the two apparently very
different moments of inertia $\mathcal{I}_{sc}, \mathcal{I}_Y$ are identical
in a perturbative treatment to leading order in $g$; this is the point
where one observes this most clearly. For the evaluation of the
expressions used here, we have used a different form of the
longitudinal projector, $(\Pi_L)^{c_2 c_1}_{i_2 i_1}
(\mathbf{x}_2,\mathbf{x}_1) = \left(\hat{\bar{\mathbf{D}}}{}^{c_2 
b_2}_{i_2}(\mathbf{x}_2) \hat{\bar{\mathbf{D}}}{}^{c_1 b_1}_{i_1}(\mathbf{x}_1)
G^{b_1 b_2}_{\Delta}(\mathbf{x}_1, \mathbf{x}_2)\right)$, where again we
don't integrate over $\mathbf{x}_1, \mathbf{x}_2$.}, thus there
is no term linear in $\rho$ that contributes to the energy, whereas the
second term simplifies to
\be
\frac{g^2}{2} \int d^3y_1\, d^3y_2\,  \rho^{d_1}(\mathbf{y}_1) G_{\Delta}^{d_1
 d_2}(\mathbf{y}_1,\mathbf{y}_2) \rho^{d_2}(\mathbf{y}_2) \qquad
 \text{with} \ - (\hat{\bar{\mathbf{D}}}_i^{ab}(\mathbf{x})
 \hat{\bar{\mathbf{D}}}_i^{bc}(\mathbf{x}))
 G_{\Delta}^{cd}(\mathbf{x}, \mathbf{y}) =
 \delta^{ad}\delta_{\mathbf{x} \mathbf{y}}.
\ee
We see that in leading order perturbation theory, the charges interact
via a potential given by the Green's function of the background
field-covariant  Laplacian. Three comments are in order: first, since
the Green's 
function depends on the background field, the variational equations
for $\bar{\mathbf{A}}$ are not independent of the charge distribution. This is
different from the electrodynamical case where radiation
(transversal terms) and static charges were completely decoupled. Second, we
see that obtaining the $\beta$ function from corrections to the
Coulomb potential (which would result for $\bar{\mathbf{A}} = 0$) is on
this
level of approximation not possible; it seems almost surprising that
we obtained the correct leading order potential, since our projection
scheme only attempts to give the correct energy to $\mathcal{O}(g^0)$.
For the first quantum corrections to the interquark potential,
we would need the energy up to $\mathcal{O}(g^4)$. Third, the fact that
the inter-charge potential is given by the Green's function of the
covariant Laplacian strongly reminds one of the 
result of  \cite{Levit:1995gm} where a rigid rotator model was
considered which was also inspired by ideas from nuclear physics.

\section{\label{sec_sum}Summary and Conclusions}
In this paper we considered the 
Kamlah expansion, which is an expansion of the expectation value of the
Hamiltonian between projected states in powers of the symmetry
generator. This is a technique successfully employed in nuclear
physics, 
but it is not clear a priori that the same can be done in a field theory with a
\textit{local} symmetry since the applicability in nuclear physics is
based on the fact that the states are strongly deformed, which comes
about usually from many particles participating in that deformed
state. \\ We tried the Kamlah expansion first for electrodynamics with
external (static) charges, where we found that the first order of the Kamlah
expansion is inconsistent with its assumptions, since at the stationary
point we find a state that is not deformed. Going to second order,
however, we found exactly the properties that we expect of a projected
energy functional, namely that it is independent of the longitudinal
parts of the $(\mathbf{\Pi}\, \mathbf{\Pi})$ correlation function and the
longitudinal part of the expectation value of $\mathbf{\Pi}$, and
the correct Coulomb energy appears in the projected energy. \\ 
This result motivated us to consider Yang-Mills theories, too, in
this framework but we restricted ourselves to the leading 
order in a perturbative expansion in powers of the coupling constant
$g$. Since we have approximated both the Gauss law operator 
and the Hamiltonian, we obtain consistency conditions from the
requirement that the approximated Hamiltonian still commutes with the
approximated Gauss law operator. Effectively, this restricts the
choice of possible background fields to those which satisfy the
(static) classical equations of motion. This condition makes the operator that
multiplies the part of the potential energy quadratic in $\mathbf{A}$
transversal w.r.t. the covariant derivative in the background
field. Since the projection subtracts off the longitudinal part (in
the generalized sense) of the 
electrical energy, we end up with an energy functional that is
independent of the longitudinal parts (in the generalized sense) of
the parameters of the Gaussian wave functional. At the stationary
point of all parameters save the 
background vector potential, the energy functional looks precisely like
the one-loop functional that one obtains in the mean field
considerations without taking into account the projector. Of course,
this is so only formally, since on the one hand the consistency
condition alluded to above ensures that the energy depends only on the
transversal degrees of freedom (thereby reducing three polarization
states to two), and, on the other hand, the longitudinal parts of the
parameters in the Gaussian wave functional are undetermined - instead
of being set to zero - just as one would expect from a projected energy
functional. We then also considered how external charges can be
included into the Kamlah expansion. Up to now, we have been
able to allow for them only to leading order in perturbation theory,
since only then the charges can be treated in a quasi-Abelian manner. \\
Three investigations seem worthwhile: \textit{first}, one has to
investigate whether the Kamlah expansion is a truly non-perturbative
expansion or whether at some stage an implicit expansion in powers of
$g$ takes place. In this context, one should also investigate further
the 'pattern' found in our discussion on electrodynamics, namely that
one seems to need the Kamlah expansion to n${}^{th}$ order to project
correctly terms containing n operators. If this would be true in
general, one would need the fourth-order Kamlah expansion for a
correct projection of the full Yang-Mills Hamiltonian. \\
\textit{Second}, one should try
to find out whether the Kamlah expansion can be carried out also to
higher orders in perturbation theory since this would
lift the restriction to background fields that satisfy the classical
equations of motion. This should be considered also in the presence of
external charges, which up to now can only be dealt with
classically. If that can be done successfully, one should face
the problem of computing the  
one-loop contribution to the inter-quark potential. If this 
reproduces the standard results, one could also try in a \textit{third}
investigation a non-perturbative
evaluation of the second-order Kamlah expansion (provided the Kamlah
expansion is truly non-perturbative and a second-order treatment is
sufficient), although this will be
a quite difficult task, since in our evaluation we relied strongly on
the factorization properties; these, however, cannot be applied so
easily in a non-perturbative framework, since
the full Gauss law operator contains products of two operators at the
same point in space. \\ 
Thus, renormalization problems, which we have
ignored completely in this paper, will have to be taken into account
in a non-perturbative manner before one can 
perform such a non-perturbative evaluation of the projected energy
functional. \\ 
Of course, when considering such non-perturbative extensions one 
should keep in mind that the restriction to Gaussian
states has tied us already to the Hartree-Fock approximation, which is
basically the two-loop approximation of the total quantum energy. 
\begin{acknowledgments}
The authors would like to thank Michael Engelhardt, Jean-Dominique
L\"ange and Markus Quandt for continuous discussion. This work was
supported by the \textit{Deutsche Forschungsgemeinschaft} under grants
DFG Mu 705/3, DFG III GK-GRK 683/1-01, DFG-Re 856/4-1 and DFG-Re 856/5-1.
\end{acknowledgments}

%
\appendix
%

\section{\label{app_conv}Conventions}
\subsection{Indices and Summation Conventions\label{AppA_index_conv}}
In this paper, four different kinds of indices appear: colour
indices, spatial indices, position indices and super-indices,
combining several of the former three kinds as will be explicitly
stated wherever they are introduced. To each kind of index a certain
range of letters is attached: 
colour indices: $a,b,c, \ldots$,
spatial indices: $i,j,k, \ldots$,
position indices: $\mathbf{x}, \mathbf{y}  \ldots$,
super-indices: $i,j,k, \ldots$\, . \\
Unless explicitly stated otherwise, Einstein's summation convention
will be used throughout. Discrete indices will be summed over,
continuous indices will be integrated over. We will also use a uniform
notation for $\delta$ functions, regardless of the nature of indices
(discrete or continuous); thus, $\delta_{\mathbf{x} \mathbf{y}}$ in this paper
corresponds to $\delta^{(3)} (\mathbf{x} - \mathbf{y})$ in the usual
notation for Dirac's $\delta$, whereas e.g. $\delta^{ab}$ is simply Kronecker's
$\delta$. \\
\subsection{Minkowski Space}
We use as metric for the Minkowski space 
\be
g_{\mu \nu} = \text{diag}(1,-1,-1,-1), 
\ee
and thus have as four-vectors $a^{\mu}$ and $a_{\mu} = g_{\mu \nu} a^{\nu}$
\be
a^{\mu} = (a_0, \mathbf{a})\hspace{0,5em}, \hspace{0.5em} a_{\mu} = (a_0,
- \mathbf{a}) 
\ee
with the scalar product
\be
a \cdot b = a_{\mu} b^{\mu} = a_0\, b_0 - \mathbf{a}.\mathbf{b}. 
\ee
Only for the four-derivative, signs are distributed differently:
\be
\partial^{\mu} = (\partial_0, - \nabla)\hspace{0,5em}, \hspace{0.5em}
\partial_{\mu} = (\partial_0, \nabla) 
\ee
with $\nabla_i = \frac{\partial}{\partial \mathbf{x}_i}$. Sometimes we
need a gradient w.r.t. another variable. This we denote by attaching
the variable name to $\nabla$, e.g. $\nabla_i^\mathbf{y} =
\frac{\partial}{\partial \mathbf{y}_i}$. 
\subsection{Group Theory Conventions}
The (hermitian) generators of the SU(N) Lie algebra satisfy the
commutation relations 
\be
~[\lambda^a, \lambda^b] = i f^{abc} \lambda^c.
\ee
The generators in
the fundamental representation(s) are normalized to (1/2):
\be
\tr{\left(\lambda^a_{\text{fund}}
\lambda^b_{\text{fund}} \right)} = 
\frac{1}{2} \delta^{ab}. 
\ee
The structure constants satisfy the \textit{Jacobi identity}
\be
f^{abr} f^{cdr} + f^{acr} f^{dbr} + f^{adr} f^{bcr} = 0
\ee
and are normalized as
\be
f^{abc} f^{dbc} = N \delta^{ad}.
\ee
\subsection{\label{App_canon_quant}Canonical Quantization}
Throughout this paper we work in the Hamiltonian framework. We 
use the Weyl gauge $A_0 = 0$, but do not fix the gauge any further. 
The usage of the Weyl gauge implies that the physical states of the
system have to satisfy the Gauss law constraint. In the absence of
external charges (which we assume throughout this appendix), the
condition reads 
\bea
&& \Gamma^a(\mathbf{x}) | \psi \rangle \stackrel{!}{=} 0 \\
\text{where} && \Gamma^a(\mathbf{x}) =
\hat{\mathbf{D}}{}^{ab}_i(\mathbf{x}) \mathbf{\Pi}^b_i(\mathbf{x}) \quad
\text{with} \quad 
\hat{\mathbf{D}}{}^{ab}_i(\mathbf{x}) = \nabla_i \delta^{ab} - f^{acb}
\mathbf{A}^c_i(\mathbf{x}). \label{eq_Gauss_law_constraint_puregauge}
\eea
$\hat{\mathbf{D}}{}^{ab}_i(\mathbf{x})$ is the covariant derivative in
the adjoint representation.
If we consider a system with external charges, the Gauss law operator
obtains additional contributions that stem from the external charges,
as is outlined in sec. \ref{sec_proj}. \\
In order to quantize the system, we do 
not have to resolve Gauss' law, since even a system with redundant
coordinates may be quantized \cite{Christ:1980ku}. We work in the
coordinate representation, i.e. $\mathbf{A}$ is realized
multiplicatively, whereas the canonical momenta are realized as
derivative operators:
\be
\langle \mathbf{A}| \mathbf{\Pi}^a_i(\mathbf{x}) | \psi \rangle =
\frac{1}{i} \frac{\delta}{\delta \mathbf{A}^a_i(\mathbf{x})} \langle
\mathbf{A} | \psi \rangle. 
\ee
This choice ensures the satisfaction of the equal-time commutation
relations automatically
\be
[ \mathbf{A}^a_i(\mathbf{x}), \mathbf{\Pi}^b_j(\mathbf{y}) ] = i
\delta^{ab} \delta_{ij} \delta_{\mathbf{x} \mathbf{y}}. \label{fund_commutator}
\ee
Factors of $g$ appear in several quantities at different places according to
the different conventions possible as will be discussed at length in
app. \ref{App_scaling}. However, contrary to some conventions used in
the literature, the meaning of $\mathbf{A}, \mathbf{\Pi}$ is always
the same irrespective of whether we choose 'perturbative' or
'non-perturbative scaling', especially no factors of $g$ enter the
commutator eq.\,(\ref{fund_commutator}). \\
The wave functionals we deal with in this paper are given in the
$\mathbf{A}$-representation
\be
\psi[\mathbf{A}] = \langle \mathbf{A} | \psi \rangle.
\ee
In the language of wave functionals the fact that 
the choice of Weyl gauge implies that $| \psi \rangle$ has to be
annihilated by the Gauss law operator is translated as
\be
\psi[\mathbf{A}^U] = \psi[\mathbf{A}]
\ee
for topologically trivial but otherwise arbitrary gauge transformations $U$.  
In this context,  $\mathbf{A}^U$ denotes the gauge transformed field
\be
(\mathbf{A}^a_i \lambda^a)^U(\mathbf{x}) = U(\mathbf{x}) \mathbf{A}^a_i(\mathbf{x}) \lambda^a U^{\dagger}(\mathbf{x}) -
i U(\mathbf{x}) \nabla_i U^{\dagger}(\mathbf{x})
\ee
with $U(\mathbf{x}) = e^{i \phi^a(\mathbf{x}) \lambda^a}$ and $\lambda^a$
are the generators of SU(N) in some representation. \\
For further details on the canonical treatment of Yang-Mills theories,
cf. \cite{Jackiw:1980ur}; more details on the Schr\"odinger picture
can be found e.g. in \cite{Jackiw:1987aq, Jackiw:1988sf}.
\subsection{Factors of g\label{App_scaling}}
In Yang-Mills theories one has basically two options concerning where
one wants to put the 
coupling constant, either in front of the action ('{\it
non-perturbative scaling}'), or in front of the 
commutator term in the field strength ('\textit{perturbative
scaling}'). In table \ref{tab_pert_nonpert_scaling} we give a short  
list concerning which
convention leads to which placing of factors of g in other quantities
of interest.
\renewcommand{\arraystretch}{1.75}
\begin{longtable*}[c]{|l|c|c|}
\hline \hline 
 &  non-perturbative scaling &  perturbative \ scaling \\ \hline
 \hline  \endhead 
  covariant derivative & $ D_{\mu} = \partial_{\mu} - i
 A_{\mu}  $  & $
 D_{\mu} = \partial_{\mu} - i \mathbf{g} A_{\mu} $ \\ \hline
  field strength &  $ F_{\mu \nu} = \partial_{\mu} A_{\nu} -
 \partial_{\nu} A_{\mu} - i [A_{\mu}, A_{\nu}]  $  &
  $ F_{\mu \nu} = \partial_{\mu} A_{\nu} - \partial_{\nu} A_{\mu} - i
 \mathbf{g}[A_{\mu}, A_{\nu}] $  \\   
 & $ = i[D_{\mu}, D_{\nu}] $ & $ =  \frac{i}{\mathbf{g}} [D_{\mu},
 D_{\nu}] $ \\  
 \hline 
  action & $ S = - \frac{1}{4 \mathbf{g}^2} F^a_{\mu \nu}(x)
 F^{a \, \mu \nu}(x) $ & $ S = - \frac{1}{4} F^a_{\mu
 \nu}(x) F^{a \,  \mu \nu}(x)$ \\ \hline
  electrical field & $ \mathbf{E}^a_{i} = F^a_{0i} $ & $ \mathbf{E}^a_{i} =
 F^a_{0i}$ \\ \hline 
  magnetic field & $  \mathbf{B}^a_{i} = -\frac{1}{2} \epsilon_{ijk} F^{a
 \, jk}  $ & $  \mathbf{B}^a_{i} = -\frac{1}{2} \epsilon_{ijk} F^{a \, jk} $ \\  
 & $ = (\nabla \times \mathbf{A})^a_i - \frac{1}{2} f^{abc}(\mathbf{A}^b
 \times \mathbf{A}^c)_i$ & $ = (\nabla \times \mathbf{A})^a_i -
 \frac{\mathbf{g}}{2} 
 f^{abc}(\mathbf{A}^b  \times \mathbf{A}^c)_i$ \\
 \hline
  momenta $\quad \mathbf{\Pi}^a_i = \frac{\partial \mathcal{L}}{\partial
 \dot{\mathbf{A}}^{a}_i} $ & $ \mathbf{\Pi}^a_i =
 \frac{1}{\mathbf{g}^2} F^a_{i0} =  -\frac{1}{\mathbf{g}^2} 
 \mathbf{E}^a_{i} $ & $ \mathbf{\Pi}^a_i =  F^a_{i0} =  -
 \mathbf{E}^a_{i} $ \\ \hline   
  Hamiltonian & $ H = \frac{\mathbf{g}^2}{2} \mathbf{\Pi}^a_i(\mathbf{x}) \mathbf{\Pi}^a_i(\mathbf{x})
 + \frac{1}{2 \mathbf{g}^2} \mathbf{B}^a_i(\mathbf{x})
 \mathbf{B}^a_i(\mathbf{x}) $ & $ H = 
 \frac{1}{2} \mathbf{\Pi}^a_i(\mathbf{x}) \mathbf{\Pi}^a_i(\mathbf{x}) + \frac{1}{2}
 \mathbf{B}^a_i(\mathbf{x}) \mathbf{B}^a_i(\mathbf{x}) $ \\  
 & $ =  \frac{1}{2 \mathbf{g}^2}( \mathbf{E}^a_i(\mathbf{x}) \mathbf{E}^a_i(\mathbf{x}) + \mathbf{B}^a_i(\mathbf{x}) \mathbf{B}^a_i(\mathbf{x})) $ & $ =
 \frac{1}{2 }( \mathbf{E}^a_i(\mathbf{x}) \mathbf{E}^a_i(\mathbf{x}) + \mathbf{B}^a_i(\mathbf{x}) \mathbf{B}^a_i(\mathbf{x})) $ \\
 \hline
  wave functional of & $ \psi[\mathbf{A}] \sim e^{-\frac{1}{\mathbf{g}^2} \mathbf{A} G^{-1} \mathbf{A}}
 $ & $  \psi[\mathbf{A}] \sim e^{- \mathbf{A} G^{-1} \mathbf{A}} $ \\ 
  'free' theory & &  \\ \hline
gauge transformations & $ U = e^{i \phi^a \lambda^a} $ & $ U = e^{i
 \mathbf{g} \phi^a  \lambda^a} $ \\ \hline 
generators & $ [(-\Gamma^a_\mathbf{x}), (-\Gamma^b_\mathbf{y})] = i
 \delta_{\mathbf{x}\mathbf{y}} f^{abc} (-\Gamma^c_\mathbf{x}) \qquad$ & $
 [\frac{1}{\mathbf{g}} (-\Gamma^a_\mathbf{x}), \frac{1}{\mathbf{g}}
 (-\Gamma^b_\mathbf{y})] =
 \frac{\mathbf{g}}{\mathbf{g}} i \delta_{\mathbf{x} \mathbf{y}} f^{abc}
 \frac{1}{\mathbf{g}} (-\Gamma^c_\mathbf{x}) \qquad $ \\ \hline
finite gauge trafos & $  e^{i \int \phi^a \Gamma^a} $ & $  
  e^{i \int \mathbf{g} \phi^a \frac{1}{\mathbf{g}} \Gamma^a} $ 
\\ 
(gluonic part) & &  \\ \hline \hline 
\caption{Placement of the coupling constant $g$ in the
'non-perturbative' and the 'perturbative' scaling
scheme. We have used the shorter form $\Gamma^a_{\mathbf{x}}$ for
$\Gamma^a(\mathbf{x})$. \label{tab_pert_nonpert_scaling}}   
\end{longtable*}
\renewcommand{\arraystretch}{0.5714}
\subsection{Generalized Projectors \label{App_proj}}
In this appendix we want to give some notions on the generalized
projectors as they are used in this paper. First let us note
that the projection operators are thought of as bilocal objects,
i.e. they depend on two spatial coordinates. Consider e.g. the (generalized)
longitudinal projector $(\Pi_L)$: 
\be
(\Pi_L)^{ab}_{ij}(\mathbf{x}, \mathbf{y}).
\ee
As is to be expected for a projector, $(\Pi_L)$ 
is idempotent (if we include an integration over the double
continuous index as well):
\be
(\Pi_L)^{ab}_{ij}(\mathbf{x}, \mathbf{y}) = \int d^3z\,
(\Pi_L)^{ac}_{ik}(\mathbf{x}, \mathbf{z})
(\Pi_L)^{cb}_{kj}(\mathbf{z}, \mathbf{y}). \label{PL_idempot} 
\ee
The (generalized) 
longitudinal projector $(\Pi_L)$ is a symmetric operator in the sense
that
\be
\int d^3x\, d^3y\, f^a_i(\mathbf{x}) (\Pi_L)^{ab}_{ij}(\mathbf{x}, \mathbf{y})
g^b_j(\mathbf{y}) = \int d^3x\, d^3y\, g^a_i(\mathbf{x})
(\Pi_L)^{ab}_{ij}(\mathbf{x}, \mathbf{y}) f^b_j(\mathbf{y}),
\label{PL_sym} 
\ee
where we have dropped boundary terms from two partial integrations.
The latter property implies that it can be diagonalized, and
from eq.\,(\ref{PL_idempot}) one can conclude that its eigenvalues are $0,1$.
The generalized
longitudinal
projector can be given in different forms; for this it might
be useful to recall that the {\it ordinary} longitudinal projector
$\PP^L$ also can be given in different forms that are equivalent under
the usual circumstances (i.e. all functions of $\mathbf{y}$ are found
to the right \footnote{In 
other words, for expressions of the type $f_i(\mathbf{x})
(\PP^L)_{ij}(\mathbf{x}, \mathbf{y}) g_j(\mathbf{x}, \mathbf{y})$
(with a sum or integral over all double indices) all these definitions are
equivalent, otherwise they will usually be not.} of
$(\PP^L)_{ij}(\mathbf{x}, \mathbf{y})$)  
\bea
(\PP^L)_{ij}(\mathbf{x}, \mathbf{y}) & = & -\Big(\nabla^\mathbf{x}_i G_{\Delta}(\mathbf{x}, \mathbf{y})\Big) \nabla^\mathbf{y}_j \label{App_ord_long_proj1} \\
& = & \Big(\nabla^\mathbf{y}_j \nabla^\mathbf{x}_i G_{\Delta}(\mathbf{x},
\mathbf{y})\Big) \label{App_ord_long_proj2} \\
& = & \int \frac{d^3k}{(2 \pi)^3} e^{i \mathbf{k}.(\mathbf{x} - \mathbf{y})}
\frac{\mathbf{k}_i \mathbf{k}_j}{\mathbf{k}^2}, \label{App_ord_long_proj3}
\eea
where $G_{\Delta}(\mathbf{x}, \mathbf{y})$ denotes the Green's function of the
Laplace operator, $- \nabla_i \nabla_i G_{\Delta}(\mathbf{x},
\mathbf{y}) = \delta_{\mathbf{x} \mathbf{y}}$. Under the given
circumstances all three definitions are equivalent; however, only the
first one really has to 'act to the right' in order to make sense. The
(ordinary) transversal projector can also easily be defined as
\be
(\PP^T)_{ij}(\mathbf{x}, \mathbf{y}) = \delta_{ij} \delta_{\mathbf{x}
\mathbf{y}} - (\PP^L)_{ij}(\mathbf{x}, \mathbf{y}).
\ee
For the generalized longitudinal projectors we can now give
corresponding definitions:
\bea
(\Pi_L)^{ab}_{ij}(\mathbf{x}, \mathbf{y}) & = & -
\Big(\hat{\bar{\mathbf{D}}}{}^{aa_1}_i 
(\mathbf{x}) G^{a_1 b_1}_{\Delta}(\mathbf{x}, \mathbf{y})\Big) \hat{\bar{\mathbf{D}}}{}^{b_1 b}_j (\mathbf{y}) \\ 
& = & \Big(\hat{\bar{\mathbf{D}}}{}^{b b_1}_j (\mathbf{y})
\hat{\bar{\mathbf{D}}}{}^{aa_1}_i (\mathbf{x}) G^{a_1
b_1}_{\Delta}(\mathbf{x}, \mathbf{y}) \Big) \\ 
& = & \sum_{n} (\xi_n^*)^a_i (\mathbf{x}) (\xi_n)^b_j(\mathbf{y})
\eea
where $G^{a_1 b_1}_{\Delta}(\mathbf{x}, \mathbf{y})$ is the Green's function
of the covariant Laplacian
$(\hat{\bar{\mathbf{D}}}\hat{\bar{\mathbf{D}}})^{c_1a_1}(\mathbf{x})$,
i.e.
$-(\hat{\bar{\mathbf{D}}}\hat{\bar{\mathbf{D}}})^{c_1a_1}(\mathbf{x})
G^{a_1 b_1}_{\Delta}(\mathbf{x}, \mathbf{y}) = \delta^{c_1
b_1}\delta_{\mathbf{x} \mathbf{y}}$, $(\xi_n)^a_i (\mathbf{x})$ are the
normalized eigenfunctions of $(\Pi_L)^{ab}_{ij}(\mathbf{x}, \mathbf{y})$
with eigenvalue 1, and we do not integrate over $\mathbf{x}, \mathbf{y}$.
We assume that the covariant Laplacian has no eigenvalues zero.
Here the corresponding generalized transversal projectors are given by 
\be
(\Pi_T)^{ab}_{ij}(\mathbf{x}, \mathbf{y}) = \delta_{ij} \delta^{ab}
\delta_{\mathbf{x} \mathbf{y}} - (\Pi_L)^{ab}_{ij}(\mathbf{x},
\mathbf{y}). \label{eq_def_Pi_T}  
\ee
%
%
%
\section{\label{app_PT}Results from Perturbation Theory}
In this appendix we want to address the following problem: assume that we start
with a wave functional that has Gaussian form. Since this wave functional is
not annihilated by the Gauss law operator $\Gamma^a(\mathbf{ x})$, we
multiply it by a polynomial. How 
do we have to fix the coefficients of the polynomial s.t. Gauss' law is
satisfied to a given order in the coupling constant g \cite{Cea:1988ku}? \\
In order to proceed, we use the same procedure as in
the main text: start out from 'non-perturbative
scaling', and then replace $\mathbf{A}$ by $\bar{\mathbf{A}} + g
\mathbf{a}$ so that the 
background field does not decouple in the perturbative limit.
The covariant derivative can then be split into a part containing the
  background field (which is completely of $\mathcal{O}(g^0)$) and a part that
  contains the fluctuation part and an explicit factor of g:
\be
\hat{\mathbf{D}}^{ab}_i(\mathbf{x}) =
\hat{\bar{\mathbf{D}}}{}^{ab}_i(\mathbf{x}) - g f^{acb} 
  \mathbf{a}^c_i(\mathbf{x}) ~\text{with}~
\hat{\bar{\mathbf{D}}}{}^{ab}_i(\mathbf{x}) = 
  \nabla_i \delta^{ab} - f^{acb} \bar{\mathbf{A}}^c_i(\mathbf{x}).
\ee
Thus the Gauss law operator reads (cf. the discussion in
sec. \ref{sec_YM_KE_2nd_order}) 
\be
\Gamma^a(\mathbf{x}) = \hat{\mathbf{D}}^{ab}_i(\mathbf{x}) \frac{1}{i}
\frac{\delta}{\delta \mathbf{a}^b_i(\mathbf{x})} =  \frac{1}{i}
\hat{\bar{\mathbf{D}}}{}^{ab}_i(\mathbf{x}) \frac{\delta}{\delta \mathbf{
a}^b_i(\mathbf{x})} - g f^{acb} \mathbf{a}^c_i(\mathbf{x})
\frac{1}{i}\frac{\delta}{\delta \mathbf{a}^b_i(\mathbf{x})}.  
\label{app_PT_def_of_Gauss}
\ee
The wave functional is of the form
\be
\Psi = P(\mathbf{a}) e^{-\frac{1}{2} \mathbf{a} \chi \mathbf{a} + i \bar{\mathbf{e}}.\mathbf{a}},
\ee
where we have used a (hopefully) obvious shorthand notation together
with the abbreviation
$\chi= \frac{1}{2} G^{-1} - 2 i \Sigma$, and  
$P(\mathbf{a})$  indicates a power series in g: 
\be
P(\mathbf{a}) = \sum_{n=0}^{\infty} g^n P^{(n)} (\mathbf{a}).
\ee 
If we now require
\be
\Gamma^a(\mathbf{x}) \Psi = 0 + \mathcal{O}(g^{N+1})
\ee
we obtain a recursion relation between the different $P^{(n)}$s:
\bea
0 = & & 
\sum_{n=0}^{N} g^n (\hat{\bar{\mathbf{D}}}{}^{ab}_i ((-\chi \mathbf{a})^b_i
+ i \bar{\mathbf{e}}^b_i) 
  P^{(n)}  -  \sum_{n=1}^{N} g^n \mathbf{a}^c_i f^{acb} ((-\chi
  \mathbf{a})^b_i + i \bar{\mathbf{e}}^b_i) P^{(n-1)} \nonumber \\ & +&
  \sum_{n=1}^{N} g^n 
  \hat{\bar{\mathbf{D}}}{}^{ab}_i \frac{\delta P^{(n)}}{\delta
\mathbf{a}^b_i} - \sum_{n=2}^{N} g^n \mathbf{a}^c_i f^{acb} \frac{\delta
      P^{(n-1)}}{\delta \mathbf{a}^b_i} = 0, \label{GaussPT1}
\eea
where it is sufficient to terminate the sums at $n=N$ since we require
Gauss' law only up to $\mathcal{O}(g^{N})$.
Note that the summation indices start at different lower values. Thus
we can study
the $\mathcal{O}(g^0)$ and $\mathcal{O}(g^1)$ cases separately\footnote{We set
  $P^{(0)} = 1$, since in the end it will be fixed by the overall
  normalization anyway; it has to be nonzero, s.t. the state does not
vanish as $g \rightarrow 0$. }:
\be
\mathcal{O}(g^0): \ \ \hat{\bar{\mathbf{D}}}{}^{ab}_i ((-\chi
\mathbf{a})^b_i + i \bar{\mathbf{e}}^b_i) = 0. 
\ee
Since this equation has to be valid for all $\mathbf{a}$, one actually
obtains three conditions
since $\hat{\bar{\mathbf{D}}}$ does
not mix real and imaginary parts:
\bea
\hat{\bar{\mathbf{D}}} \chi & = & 0 \  \to
\hat{\bar{\mathbf{D}}}{}^{ab}_i(\mathbf{x}) 
(G^{-1})^{ab}_{ij}(\mathbf{x}, \mathbf{y})   =  0 \ \text{and} \
\hat{\bar{\mathbf{D}}}^{ab}_i(\mathbf{x}) \Sigma^{ab}_{ij}(\mathbf{x},
\mathbf{y})   =  0 \label{GaussPTOg0a} \\  
\hat{\bar{\mathbf{D}}} \bar{\mathbf{e}} & = & 0,
\  \text{i.e.} \hat{\bar{\mathbf{D}}}{}^{ab}_i(\mathbf{x})
\bar{\mathbf{e}}^{b}_i(\mathbf{x})  =  0. \label{GaussPTOg0b}  
\eea
Note the similarity with the QED condition of having a purely
transversal kernel. \\
Carrying this procedure further to higher orders in $g$ will lead to
certain difficulties, e.g. that $\psi$ will be annihilated only if one
takes $\mathbf{a}$ to stem from a subspace satisfying the background
field condition $\hat{\bar{\mathbf{D}}} \mathbf{a} = 0$, thereby
invalidating the whole concept, since our states should be such that
Gauss' law is satisfied for all $\mathbf{a}$ without further gauge
fixing. However, to $\mathcal{O}(g^0)$, the concept works, and is
compatible with the results from the Kamlah expansion as seen in
sec. \ref{sec_YMT}. \\
The main lesson from this section is that a pure Gaussian state can satisfy
Gauss' law only to order $g^0$. This is the reason why we can
formulate an exactly gauge invariant Gaussian ground state for electrodynamics.
%
%
%
\section{Derivations \label{app_deriv}}
This appendix contains some of the technical details of the different
derivations.
\subsection{Electrodynamics, Expansion to Second Order
\label{app_elec_kam_2nd}} 
\subsubsection{Generalities}
Here one starts from the expression
(cf. eqs.\,(\ref{Abelian_Kamlah_exp1}, \ref{Abelian_Kamlah_exp2}))  
\be
\begin{array}{lcl}
\langle H e^{i \int \phi (\Gamma - \rho)} \rangle & = &
\hspace*{0.8em} A_0 \langle e^{i \int 
  \phi (\Gamma - \rho)} \rangle  + A_1(\mathbf{y}) \langle \{
  \Gamma(\mathbf{y}) - \langle \Gamma(\mathbf{y}) \rangle \} \
  e^{i \int \phi (\Gamma - \rho)} \rangle  
\vspace*{0.3em} \\ & &  +  A_2(\mathbf{y},\mathbf{z}) \langle
  \{\Gamma(\mathbf{y}) - 
\langle \Gamma(\mathbf{y}) \rangle \} \{\Gamma(\mathbf{z}) -\langle
  \Gamma(\mathbf{z}) \rangle \} \ e^{i \int \phi (\Gamma - \rho)}
  \rangle. 
\end{array}
\label{definingsecondorderphoton}
\ee
The coefficients will be determined as follows: one needs three
equations, which are obtained by 
\begin{itemize}
\item[(a)] setting $\phi=0$, 
\item[(b)] performing a functional 
derivative w.r.t. $\phi(\mathbf{x}_1)$ and setting $\phi=0$ afterwards,
\item[(c)] performing two functional derivatives, w.r.t. $\phi(\mathbf{x}_1),
\phi(\mathbf{x}_2)$ and setting $\phi=0$ afterwards. 
\end{itemize}
After a bit of manipulation, one obtains 
\bea
A_0 & = & \langle H \rangle - A_2(\mathbf{y},\mathbf{z})
\langle \Delta(\mathbf{y}) \Delta(\mathbf{z}) \rangle, \\
A_1(\mathbf{y}) & = & \left( \langle H \Delta(\mathbf{x}_1)
\rangle -  A_2(\mathbf{y}_1,\mathbf{z}_1) \langle
\Delta(\mathbf{y_1}) \Delta({\mathbf{z}_1}) \Delta({\mathbf{x}_1})
\rangle\right) \Theta^{-1}(\mathbf{x}_1,\mathbf{y})  
\eea 
and for $A_2$ one obtains an even less pleasant equation 
\bea
& & \langle H \Delta({\mathbf{x}_1}) \Delta({\mathbf{x}_2}) \rangle -
\langle H \rangle \langle \Delta({\mathbf{x}_1})  \Delta({\mathbf{x}_2}) \rangle  
- \langle H \Delta(\mathbf{z}) \rangle \Theta^{-1}(\mathbf{z}, \mathbf{y})
  \langle \Delta(\mathbf{y}) \Delta(\mathbf{x}_1) \Delta(\mathbf{x}_2) \rangle
\nonumber \\  
& = & 
A_2(\mathbf{y}, \mathbf{z}) \Big[\langle \Delta(\mathbf{y}) \Delta(\mathbf{z})
\Delta({\mathbf{x}_1})  \Delta({\mathbf{x}_2}) \rangle -
  \langle \Delta(\mathbf{y}) \Delta(\mathbf{z}) \rangle  \langle
\Delta({\mathbf{x}_1}) \Delta({\mathbf{x}_2}) \rangle 
 \nonumber \\ & & \hspace*{4em} -  
  \langle \Delta(\mathbf{y}) \Delta(\mathbf{z}) \Delta(\mathbf{y}_1) \rangle
\Theta^{-1}(\mathbf{y}_1, \mathbf{y}_2)
  \langle \Delta(\mathbf{y_2}) \Delta({\mathbf{x}_1}) \Delta({\mathbf{x}_2})
  \rangle \Big],
\label{A2detphoton}
\eea
with $\Delta(\mathbf{x}) = \Gamma (\mathbf{x}) - \langle \Gamma
(\mathbf{x}) \rangle$ and $\Theta^{-1}$ is defined via
$\Theta^{-1}(\mathbf{x}, \mathbf{y}) \langle \Delta(\mathbf{y})
\Delta(\mathbf{z})  \rangle = \delta_{\mathbf{x} \mathbf{z}}$. 
Eq.\,(\ref{A2detphoton}) simplifies significantly if one exploits some
of the factorization features of Gaussian states, 
\bea
\langle \Delta(\mathbf{x}) \Delta(\mathbf{y}) \Delta(\mathbf{z})  \rangle &=& 0\label{AbelianGaussfakt0}  \\
\langle \Delta(\mathbf{y}) \Delta(\mathbf{z}) \Delta({\mathbf{x}_1})
\Delta({\mathbf{x}_2}) \rangle & = & \langle 
\Delta(\mathbf{y}) \Delta(\mathbf{z}) \rangle \langle \Delta({\mathbf{x}_1})
\Delta({\mathbf{x}_2}) \rangle + \langle
\Delta({\mathbf{x}_1}) \Delta(\mathbf{z}) \rangle \langle \Delta(\mathbf{y})
\Delta({\mathbf{x}_2}) \rangle \nonumber \\ & & + \langle
\Delta({\mathbf{x}_2}) \Delta(\mathbf{z}) \rangle \langle \Delta({\mathbf{
x}_1}) \Delta(\mathbf{y}) \rangle. 
\label{AbelianGaussfakt} 
\eea
This allows to rewrite eq.\,(\ref{A2detphoton}) as
\bea
& & \langle H \Delta({\mathbf{x}_1}) \Delta({\mathbf{x}_2}) \rangle -
 \langle H  \rangle \langle \Delta({\mathbf{x}_1})
 \Delta({\mathbf{x}_2}) \rangle \nonumber  \\ &=&  
A_2(\mathbf{y},\mathbf{z}) \left(\langle \Delta(\mathbf{y})
 \Delta({\mathbf{x}_1}) 
\rangle \langle \Delta(\mathbf{z}) 
  \Delta({\mathbf{x}_2}) \rangle + \langle \Delta(\mathbf{y}) \Delta({\mathbf{
  x}_2}) \rangle  \langle \Delta(\mathbf{z}) \Delta({\mathbf{x}_1}) \rangle
  \right). 
\eea
In order to obtain $A_2$ explicitly, one can again use $\Theta^{-1}$: 
\bea & & 
\Big( \langle \Delta(\mathbf{y}) \Delta({\mathbf{x}_1}) \rangle \langle
\Delta(\mathbf{z}) \Delta({\mathbf{x}_2}) \rangle + \langle \Delta(\mathbf{y})
\Delta({\mathbf{x}_2}) \rangle  \langle \Delta(\mathbf{z}) \Delta({\mathbf{
x}_1}) \rangle \Big) \frac{1}{2} \Theta^{-1} (\mathbf{x}_1,\mathbf{y}_1)
\Theta^{-1} (\mathbf{x}_2,\mathbf{y}_2) \nonumber \\ &=& \frac{1}{2} 
(\delta_{\mathbf{y} \mathbf{y}_1} \delta_{\mathbf{z} \mathbf{y}_2} +
\delta_{\mathbf{y} \mathbf{y}_2} \delta_{\mathbf{z} \mathbf{y}_1}),   
\eea
giving (since $A_2(\mathbf{y}_1, \mathbf{y}_2) = A_2(\mathbf{y}_2,
\mathbf{y}_1)$) 
\be
A_2(\mathbf{y}_1, \mathbf{y}_2) =  \Big(\langle H \Delta({\mathbf{x}_1})
 \Delta({\mathbf{x}_2}) \rangle -  \langle H  \rangle \langle
 \Delta({\mathbf{x}_1}) \Delta({\mathbf{x}_2}) \rangle\Big)
\frac{1}{2} \Theta^{-1} 
 (\mathbf{x}_1,\mathbf{y}_1) \Theta^{-1} (\mathbf{x}_2,\mathbf{y}_2).
\ee
This result leads directly to eq.\,(\ref{sec_order_en_func_ED}) in the
main text. 
\subsubsection{Explicit Form of Energy Functional}
 We will
now turn to the evaluation of the energy. In addition to the terms we
have computed in the context of the first order Kamlah expansion, we need 
to compute $\langle \mathbf{B}_i(\mathbf{x}) \mathbf{B}_j(\mathbf{y})
\Gamma(\mathbf{z}_1) 
\Gamma(\mathbf{z}_2) \rangle$. If we
again - as in the first order calculation - take either $\Sigma = 0$
or $G^{-1}$ and $\Sigma$ translationally invariant, the matrix element
factorizes, 
\be
\langle \mathbf{B}_i(\mathbf{x}) \mathbf{B}_j(\mathbf{y}) \Gamma(\mathbf{z}_1)
\Gamma(\mathbf{z}_2) 
\rangle = \langle \mathbf{B}_i(\mathbf{x}) \mathbf{B}_j(\mathbf{y}) \rangle \langle
\Gamma(\mathbf{z}_1) \Gamma(\mathbf{z}_2) \rangle 
\ee
which is again plausible, since there should be no corrections to the
magnetic part of the energy as it depends on the transversal degrees of
freedom only. The next ingredient we have to compute for $\langle \Delta H
\Delta({\mathbf{x}_1}) \Delta({\mathbf{x}_2}) \rangle$ 
is the electric energy, which gives a rather simple
contribution, too. We can evaluate it using 
the same factorization properties that underly
eq.\,(\ref{AbelianGaussfakt}), and obtain altogether:
\be
\langle \Delta H \Delta({\mathbf{x}_1}) \Delta({\mathbf{x}_2}) \rangle =
\langle \Gamma(\mathbf{x}_2)
\mathbf{\Pi}_i(\mathbf{x}) \rangle_c  \langle \Gamma(\mathbf{x}_1) \mathbf{
\Pi}_i(\mathbf{x}) \rangle_c.  
\ee
This expression enters the correction of the mean-field energy 
\be
\frac{1}{2} \langle \Delta H \Delta({\mathbf{x}_1}) \Delta({\mathbf{x}_2}) \rangle
\Theta^{-1}(\mathbf{x}_1,\mathbf{x}_2) =  \frac{1}{2} \langle
\PP^{L}_{ij}(\mathbf{x}_1, \mathbf{x}_2) 
\mathbf{\Pi}_j(\mathbf{x}_2) \mathbf{\Pi}_i(\mathbf{x}_1) \rangle_c, 
\ee
where $\PP^{L}_{ij}(\mathbf{x}_2, \mathbf{x}_1)$ denotes the longitudinal
projector as defined \footnote{In this context there is no problem with
the other definitions eqs.\,(\ref{App_ord_long_proj2},
\ref{App_ord_long_proj3}).} e.g. in eq.\,(\ref{App_ord_long_proj1}); 
translational invariance, that has been assumed before in the first
order calculation, implies that the
electrical energy can be written as 
\be
\frac{1}{2} \langle  \mathbf{\Pi}_i(\mathbf{x}) \mathbf{\Pi}_i(\mathbf{x}) \rangle =
\frac{1}{2} \langle \PP^{L}_{ji} (\mathbf{y}, \mathbf{x}) 
\mathbf{\Pi}_i(\mathbf{x}) \mathbf{\Pi}_j(\mathbf{y}) \rangle + 
\frac{1}{2} \langle \PP^{T}_{ji} (\mathbf{y}, \mathbf{x})
\mathbf{\Pi}_i(\mathbf{x}) \mathbf{\Pi}_j(\mathbf{y}) \rangle    
\ee
and thus the second order Kamlah expansion precisely subtracts off the
longitudinal (gauge variant) part of the kernel.
The other term needed to compute $\left. \langle H \PP^{\rho} \rangle/\langle
\PP^{\rho} \rangle \second$ is even simpler,
\be
\frac{1}{2} \langle \Delta H \Delta({\mathbf{x}_1}) \Delta({\mathbf{x}_2}) \rangle
\Theta^{-1}(\mathbf{x}_1,\mathbf{y}_1) \Theta^{-1}(\mathbf{x}_2,\mathbf{y}_2) =
G_{\Delta}(\mathbf{y}_1, \mathbf{y}_2). 
\ee
Similar to the first-order discussion, we subtract off the part $\langle
\mathbf{\Pi}_i^L \rangle \langle \mathbf{\Pi}_i^L \rangle$ (thus completing
$ \langle \PP^{L}_{ji} 
\mathbf{\Pi}_i \mathbf{\Pi}_j \rangle_c$ to $ \langle \PP^{L}_{ji} \mathbf{
\Pi}_i \mathbf{\Pi}_j \rangle$) and add it back 
in afterwards. Thereby, we obtain eq.\,(\ref{ED_proj_energ}).
\subsection{Yang-Mills Theories, Expansion to Second Order
\label{app_YM_kam_2nd}} 
\subsubsection{Generalities}
We follow the same procedure as it was outlined in the case of
electrodynamics in app.\,\ref{app_elec_kam_2nd}.
From eq.\,(\ref{YMKamlah2ndansatz}) we obtain by this procedure again
three equations: 
\bea
\langle H \rangle &=& A_0 + A^{ab}_2(\mathbf{y},\mathbf{
z}) \langle \Delta^a(\mathbf{y}) \Delta^b(\mathbf{z}) \rangle, \\
\langle H \mathcal{G}^c(\mathbf{x}) \rangle &=& A_0 \langle\mathcal{G}^c(\mathbf{
x}) \rangle 
 + A_1^a(\mathbf{y}) \langle \Delta^a(\mathbf{y}) \mathcal{G}^c(\mathbf{
x}) \rangle \nonumber \\
& & \hspace*{4.6em} + A^{ab}_2(\mathbf{y},\mathbf{z}) \langle
\Delta^a(\mathbf{y}) \Delta^b(\mathbf{z}) \mathcal{G}^c(\mathbf{x}) \rangle 
\\
\langle H \frac{1}{2} \{ \mathcal{G}^{c_1}(\mathbf{x}_1), {\cal
 G}^{c_2}(\mathbf{x}_2) \} 
\rangle &=& A_0 
\langle \frac{1}{2} \{ \mathcal{G}^{c_1}(\mathbf{x}_1), \mathcal{G}^{c_2}(\mathbf{
x}_2) \} \rangle + A_1^a(\mathbf{y}) \langle \Delta^a(\mathbf{y})
\frac{1}{2} \{ \mathcal{
G}^{c_1}(\mathbf{x}_1), \mathcal{G}^{c_2}(\mathbf{x}_2) \}   \rangle
\nonumber \\ & &  
+ A^{ab}_2(\mathbf{y},\mathbf{z}) \langle
\Delta^a(\mathbf{y}) \Delta^b(\mathbf{z}) \frac{1}{2} \{\mathcal{G}^{c_1}(\mathbf{
x}_1), \mathcal{G}^{c_2}(\mathbf{x}_2) \}
\rangle \label{YMKamlah2nd_1}
\eea
where in the last equation, eq.\,(\ref{YMKamlah2nd_1}), we have used
the abbreviation $\{ \mathcal{G}^{c_1}(\mathbf{x}_1), \mathcal{G}^{c_2}(\mathbf{x}_2) \} =
\mathcal{G}^{c_1}(\mathbf{x}_1) \mathcal{G}^{c_2}(\mathbf{x}_2) + {\cal
G}^{c_2}(\mathbf{x}_2) \mathcal{G}^{c_1}(\mathbf{x}_1)$. We then can - again -
express $A_0$, $A_1$ in terms of $A_2$ and unprojected expectation
values; the equation for 
$A_2$ is very similar to eq.\,(\ref{A2detphoton}) in the electrodynamic
calculation, but here it is even worse, since $[\Delta^a(\mathbf{x}),
\Delta^b(\mathbf{y})] \neq 0$. 
In the case of electrodynamics, significant progress could be made
using the factorization features eqs.\,(\ref{AbelianGaussfakt0},
\ref{AbelianGaussfakt}). In order to mimic the treatment presented
there, and thus to make further progress, we have to resort to the
perturbative expansion of eq.\,(\ref{YMKamlah2ndansatz}) as outlined
in sec. \ref{sec_YM_KE_2nd_order}. 
\subsubsection{Factorizations}
For the quasi-Abelian form $\bar{\Gamma}^a$ we again have the factorization
properties of Gaussian states similar to
eqs.\,(\ref{AbelianGaussfakt0}, \ref{AbelianGaussfakt}):
\bea
\langle \bar{\Delta}^a(\mathbf{x}) \bar{\Delta}^b(\mathbf{y})
\bar{\Delta}^c(\mathbf{z})  \rangle =& & 0, \\ 
\langle \bar{\Delta}^a(\mathbf{y}) \bar{\Delta}^b(\mathbf{z})
\bar{\Delta}^{c_1}(\mathbf{x}_1) \bar{\Delta}^{c_2}(\mathbf{x}_2) 
\rangle = & & \langle 
\bar{\Delta}^a(\mathbf{y}) \bar{\Delta}^b(\mathbf{z}) \rangle \langle
\bar{\Delta}^{c_1}(\mathbf{x}_1) 
\bar{\Delta}^{c_2}(\mathbf{x}_2) \rangle \nonumber \\ & + & \langle 
\bar{\Delta}^{c_1}(\mathbf{x}_1) \bar{\Delta}^b(\mathbf{z}) \rangle \langle
\bar{\Delta}^a(\mathbf{y}) 
\bar{\Delta}^{c_2}(\mathbf{x}_2) \rangle \nonumber \\ & + &  \langle 
\bar{\Delta}^{c_2}(\mathbf{x}_2) \bar{\Delta}^b(\mathbf{z}) \rangle \langle
\bar{\Delta}^{c_1}(\mathbf{x}_1) \bar{\Delta}^a(\mathbf{y}) \rangle,
\label{PT0Gaussfakt} 
\eea
where we have introduced the notation $\bar{\Delta}^a(\mathbf{x}) =
\bar{\Gamma}^a(\mathbf{x}) - \langle \bar{\Gamma}^a(\mathbf{x}) \rangle$. 
\subsubsection{Derivation \label{app_YM_kam_2nd_detail}}
We now use the factorization properties to compute $A_2$ (or, if we
hadn't required $A_2$ to be symmetric from the outset, the symmetric
part of $A_2$; but since the $\bar{\Gamma}$s commute anyway, the
distinction does not play a role):
\bea
A_2^{d_1 d_2}(\mathbf{y}_1, \mathbf{y}_2) &=& \frac{1}{2}
\left[\langle H  \bar{\Delta}^{c_1}(\mathbf{x}_1)
\bar{\Delta}^{c_2}(\mathbf{x}_2)  \rangle - \langle H 
\rangle \langle \bar{\Delta}^{c_1}(\mathbf{x}_1)  \bar{\Delta}^{c_2}(\mathbf{x}_2)
\rangle \right] \times \nonumber \\ & & \quad (\Theta^{-1})^{c_1
d_1}(\mathbf{x}_1, \mathbf{y}_1) 
(\Theta^{-1})^{c_2 d_2}(\mathbf{x}_2, \mathbf{y}_2). 
\eea
With this, we can give now explicitly the projected, normalized energy
functional to second order in the Kamlah expansion (in absence of
external charges):
\bea
\left. \frac{\langle H \PP \rangle}{\langle \PP \rangle} \second
&=& \langle H 
\rangle - \frac{1}{2} \langle \Delta H
\bar{\Delta}^{c_1}(\mathbf{x}_1) 
\bar{\Delta}^{c_2}(\mathbf{x}_2) \rangle (\Theta^{-1})^{c_2 c_1}(\mathbf{
x}_2,\mathbf{x}_1) \nonumber \\ & & \hspace*{1.85em} 
-  \langle \Delta H \bar{\Delta}^{c_1}(\mathbf{x}_1) 
 \rangle (\Theta^{-1})^{c_1 c_2}(\mathbf{x}_1,\mathbf{x}_2) \langle
\bar{\Gamma}^{c_2}(\mathbf{x}_2) \rangle \label{YM_Kamlah_final1} \\ & &
\hspace*{1.85em}  + 
\frac{1}{2} \langle \Delta H
 \bar{\Delta}^{c_1}(\mathbf{x}_1) 
\bar{\Delta}^{c_2}(\mathbf{x}_2) \rangle (\Theta^{-1})^{c_1 d_1}(\mathbf{
x}_1,\mathbf{y}_1)  (\Theta^{-1})^{c_2 d_2}(\mathbf{x}_2,\mathbf{y}_2) \langle 
\bar{\Gamma}^{d_1}(\mathbf{y}_1) \rangle \langle \bar{\Gamma}^{d_2}(\mathbf{y}_2) 
 \rangle. \nonumber
\eea
As in the case of electrodynamics, we may here also insert
$\bar{\Gamma}$ into the Kamlah expansion expression
instead of $H$. This is
due to the fact that $\bar{\Gamma}^a$ is gauge-invariant w.r.t. gauge
transformations carried out by $e^{i \int \varphi^a \bar{\Gamma}^a}$.
If we do this, the terms analogous to the second and the fourth term in
eq.\,(\ref{YM_Kamlah_final1}) (with $H$ replaced by $\bar{\Gamma}^a$)
are zero, and the terms analogous to the first and third 
term cancel, so that we obtain the correct result $\left. \langle
\bar{\Gamma}^a \PP \rangle\secondt = 0$. The computation can also be carried
out for $\left. \langle \bar{\Gamma}^a \bar{\Gamma}^b \PP \rangle/\langle \PP
\rangle \secondt$, and one obtains also the correct result, namely $0$; the
term analogous to the second
term gives $-\langle \bar{\Delta}^a \bar{\Delta}^b \rangle$, whereas
the terms analogous to the third and fourth term add up to give $-
\langle \bar{\Gamma}^a \rangle 
\langle \bar{\Gamma}^b \rangle$, thus cancelling (together with the
term analogous to the second) the term analogous to the first term. \\ 
In order to be able to compute the projected energy functional to
$\mathcal{O}(g^0)$, we now need the following ingredients:
\bea
 \langle  \Delta(\mathbf{\Pi}^a_i(\mathbf{x}) \mathbf{\Pi}^a_i(\mathbf{x}))
 \bar{\Delta}^{c_1}(\mathbf{x}_1) \rangle & 
 = & 2 \langle  \mathbf{\Pi}^a_i(\mathbf{x})
 \bar{\Delta}^{c_1}(\mathbf{x}_1)  \rangle \langle \mathbf{\Pi}^a_i(\mathbf{x})
 \rangle,  \\ 
 \langle \Delta(\mathbf{\Pi}^a_i(\mathbf{x}) \mathbf{\Pi}^a_i(\mathbf{
 x})) \bar{\Delta}^{c_1}(\mathbf{x}_1) \bar{\Delta}^{c_2}(\mathbf{x}_2)
 \rangle & = & 2  \langle \mathbf{\Pi}^a_i(\mathbf{x})
 \bar{\Delta}^{c_1}(\mathbf{x}_1) \rangle \langle \mathbf{\Pi}^a_i(\mathbf{x})
 \bar{\Delta}^{c_2}(\mathbf{x}_2) \rangle, \hspace*{2em} \\   
 \langle \Delta(\mathbf{B}^a_i(\mathbf{x}) \mathbf{B}^a_i(\mathbf{x}))
 \bar{\Delta}^{c_1}(\mathbf{x}_1) \rangle & = & 0, \\  
 \langle \mathbf{B}^a_i(\mathbf{x}) \mathbf{B}^a_i(\mathbf{x})
\bar{\Gamma}^{c_1}(\mathbf{x}_1) \bar{\Gamma}^{c_2}(\mathbf{x}_2)  \rangle &
= &  - \frac{1}{2} \int d^3x\, \hat{\bar{\mathbf{D}}}{}^{c_1
 d_1}_{l_1}(\mathbf{x}_1) \hat{\bar{\mathbf{D}}}{}^{c_2 d_2}_{l_2}(\mathbf{x}_2)
 M^{d_1 d_2}_{l_1 l_2}(\mathbf{x};\mathbf{x}_1,\mathbf{x}_2)  \nonumber \\  
 & = & 0, \label{BBGammaGammaEqualZero}
\eea
where we have set $\Sigma$ to zero here and in the remaining computations
of this appendix. Also the expression $\int d^3x\,
\hat{\bar{\mathbf{D}}}{}^{c_1 d_1}_{l_1}(\mathbf{x}_1)
\hat{\bar{\mathbf{D}}}{}^{c_2 d_2}_{l_2}(\mathbf{x}_2) 
 M^{d_1 d_2}_{l_1 l_2}(\mathbf{x};\mathbf{x}_1,\mathbf{x}_2)$ requires a bit of
special attention: there is no integration over $\mathbf{x}_1,\mathbf{x}_2$
although they appear more than once, but we have to integrate over
$\mathbf{x}$, although it seems to appear only once (but this is an
artifact of the notation introduced in eq.\,(\ref{MFexpansion})).  
The last identity eq.\,(\ref{BBGammaGammaEqualZero}) is true since we
had to restrict the allowed 
background fields to fulfil the classical equations as discussed above.
In addition, we need a simple expression for $\Theta^{-1}$; in the
electrodynamics calculation we saw that we had to assume translational
invariance of $G, \Sigma$. Here we'll need something similar, namely
we require
\bea
(G^{-1})^{ab}_{ij}(\mathbf{x},\mathbf{y}) &\stackrel{!}{=}&
\hspace*{0.77em} (\Pi_L)^{a c_1}_{i 
i_1}(\mathbf{x},\mathbf{x}_1)  (G^{-1})^{c_1 c_2}_{i_1 i_2}(\mathbf{x}_1,\mathbf{
x}_2) (\Pi_L)^{c_2 
b}_{i_2 j}(\mathbf{x}_2,\mathbf{y}) \nonumber \\ & & + (\Pi_T)^{a 
c_1}_{i i_1}(\mathbf{x},\mathbf{x}_1)  (G^{-1})^{c_1 c_2}_{i_1 i_2}(\mathbf{
x}_1,\mathbf{x}_2) (\Pi_T)^{c_2 
b}_{i_2 j}(\mathbf{x}_2,\mathbf{y}) \nonumber \\
&& =  (G^{-1}_{LL})^{ab}_{ij}(\mathbf{x},\mathbf{y})
+(G^{-1}_{TT})^{ab}_{ij}(\mathbf{x},\mathbf{y}) \label{restr_on_G_YM} 
\eea
with the (generalized) longitudinal projector\footnote{Some properties
of generalized projectors are given in appendix\,\ref{App_proj}. }
\be
(\Pi_L)^{a b}_{i j}(\mathbf{x},\mathbf{y}) = -\hat{\bar{\mathbf{D}}}{}^{a
c}_{i}(\mathbf{x}) 
G_{\Delta}^{cd} (\mathbf{x},\mathbf{y}) \hat{\bar{\mathbf{D}}}{}^{db}_{j}(\mathbf{y})
\hspace*{0.25em}, \hspace*{2.75em} \text{where}~-(\hat{\bar{\mathbf{
D}}}{}\hat{\bar{\mathbf{D}}}{})^{ac}(\mathbf{x}) G_{\Delta}^{cb} (\mathbf{x},\mathbf{
y}) = \delta^{ab} \delta_{\mathbf{x}\mathbf{y}},
\ee       
(no integration over $\mathbf{x}, \mathbf{y}$) 
and the (generalized) transversal projector $(\Pi_T)^{a b}_{i
j}(\mathbf{x},\mathbf{y}) = \delta^{ab} \delta_{ij}
\delta_{\mathbf{x}\mathbf{y}} - (\Pi_L)^{a b}_{i j}(\mathbf{x},\mathbf{y})$. 
Thus, we don't want to allow
for non-zero cross terms $\Pi_L G^{-1} \Pi_T, \Pi_T G^{-1}
\Pi_L$. For the explicit construction of $\Theta^{-1}$ this has the
advantage that one can invert the longitudinal and transversal parts
individually:
\bea
& & (G^{-1}_{LL})^{ab}_{ij}(\mathbf{x},\mathbf{y}) (G_{LL})^{bc}_{jk}(\mathbf{
y},\mathbf{z}) = (\Pi_L)^{a c}_{i k}(\mathbf{x},\mathbf{z}) 
\\ & \text{and} &    
(G^{-1}_{TT})^{ab}_{ij}(\mathbf{x},\mathbf{y}) (G_{TT})^{bc}_{jk}(\mathbf{
y},\mathbf{z}) = (\Pi_T)^{a c}_{i k}(\mathbf{x},\mathbf{z}).
\eea
We now want to relate the inverse of $\hat{\bar{\mathbf{D}}}{}_{i}
\hat{\bar{\mathbf{D}}}{}_{j} (G^{-1})$ to $\hat{\bar{\mathbf{D}}}{}_{i}
\hat{\bar{\mathbf{D}}}{}_{j} G$:
we start with the observation that, since $G^{-1}_{LL} = \Pi_L G^{-1}
\Pi_L$, we can write (no integration over $\mathbf{x},\mathbf{z}$)
\be
(G^{-1}_{LL})_{ij}^{ab}(\mathbf{x},\mathbf{z}) = \hat{\bar{\mathbf{D}}}{}^{a
c_1}_{i}(\mathbf{x}) 
\hat{\bar{\mathbf{D}}}{}^{b c_2}_{j}(\mathbf{z}) (H_L)^{c_1 c_2}(\mathbf{x},\mathbf{z})    
\ee
and correspondingly (again no integration over $\mathbf{x},\mathbf{z}$)
\be
(G_{LL})_{ij}^{ab}(\mathbf{x},\mathbf{z}) = \hat{\bar{\mathbf{D}}}{}^{a c_1}_{i}(\mathbf{x})
\hat{\bar{\mathbf{D}}}{}^{b c_2}_{j}(\mathbf{z}) (K_L)^{c_1 c_2}(\mathbf{x},\mathbf{z}),    
\ee
where $H_L, K_L$ are auxiliary functions that have the important
characteristic that they \textit{don't carry spatial indices}.
From the requirement $\int d^3y\, (G^{-1}_{LL})_{ij}^{ab}(\mathbf{x},\mathbf{y})
(G_{LL})_{jk}^{bc}(\mathbf{y},\mathbf{z}) = (\Pi_L)^{ac}_{ik}(\mathbf{x},\mathbf{
z})$ we obtain 
\be
G_{\Delta}^{a_1 c_1} (\mathbf{x},\mathbf{y}) = \int d^3z\,
\left((\hat{\bar{\mathbf{D}}}{}\hat{\bar{\mathbf{D}}}{})^{b_2 b_1}(\mathbf{z})
(H_L)^{a_1 b_1}(\mathbf{x},\mathbf{z}) 
\right) (K_L)^{b_2 c_1}(\mathbf{z},\mathbf{y}). \label{det_K}
\ee
If on the other hand we start from the defining equation of
$\Theta^{-1}$
\be
\int d^3y \,
\left(\frac{1}{4} \hat{\bar{\mathbf{D}}}{}^{a b_1}_{i}(\mathbf{x})
\hat{\bar{\mathbf{D}}}{}^{b b_2}_{j}(\mathbf{y}) (G^{-1})_{ij}^{b_1 b_2}(\mathbf{
x},\mathbf{y}) \right) 
(\Theta^{-1})^{bc}(\mathbf{y},\mathbf{z}) = \delta^{ac} \delta_{\mathbf{x}\mathbf{z}}
\ee
we end up with 
\be
4 G_{\Delta}^{a_1 c_1} (\mathbf{x},\mathbf{y}) = \int d^3z\,
\left((\hat{\bar{\mathbf{D}}}{}\hat{\bar{\mathbf{D}}}{})^{b_2 b_1}(\mathbf{z})
(H_L)^{a_1 b_1}(\mathbf{x},\mathbf{z}) 
\right) (\Theta^{-1})^{b_2 c_1}(\mathbf{z},\mathbf{y}). \label{det_Theta}
\ee
From the comparison of eq.\,(\ref{det_Theta}) with eq.\,(\ref{det_K}),
we obtain
\be
(\Theta^{-1})^{b_2 c_1}(\mathbf{z},\mathbf{y}) = 4 (K_L)^{b_2 c_1}(\mathbf{z},\mathbf{y})
\ee
and thus conclude 
\be  
\hat{\bar{\mathbf{D}}}{}^{b b_1}_{i}(\mathbf{x}) \hat{\bar{\mathbf{D}}}{}^{c
b_2}_{j}(\mathbf{y}) 
(\Theta^{-1})^{b_1 b_2}(\mathbf{x},\mathbf{y}) = 4  (G_{LL})_{ij}^{bc}(\mathbf{
x},\mathbf{y}), 
\ee
where again we don't integrate over $\mathbf{x}, \mathbf{y}$.
We can now compute the correction terms to the Kamlah expansion very
easily, using the fact that e.g. $\langle \mathbf{\Pi}^a_i(\mathbf{x})
\bar{\Delta}^b(\mathbf{y}) \rangle =
\hat{\bar{\mathbf{D}}}{}^{bc}_j(\mathbf{y}) 
(G^{-1}_{LL})^{ac}_{ij}(\mathbf{x},\mathbf{y})$: 
\bea \hspace*{-2em}
\begin{array}{llll}
1. & & \langle \Delta (\mathbf{\Pi}^a_i(\mathbf{x}) \mathbf{
\Pi}^a_i(\mathbf{x})) 
\bar{\Delta}^{c_1} (\mathbf{x}_1) \bar{\Delta}^{c_2}(\mathbf{x}_2) \rangle
(\Theta^{-1})^{c_1 c_2}(\mathbf{x}_1, \mathbf{x}_2) \\ & = & 2 (\mathbf{
\Pi}_L)^{c_2 c_1}_{i_2i_1} (\mathbf{x}_2,\mathbf{x}_1) \langle \mathbf{
\Pi}^{c_1}_{i_1}(\mathbf{x}_1) \mathbf{\Pi}^{c_2}_{i_2}(\mathbf{x}_2) \rangle_c \\
2. & & \langle \mathbf{\Pi}^a_i(\mathbf{x}) \mathbf{\Pi}^a_i(\mathbf{x})
\bar{\Delta}^{c_1} (\mathbf{x}_1)  \rangle  (\Theta^{-1})^{c_1
c_2}(\mathbf{x}_1, \mathbf{x}_2) \langle \bar{\Gamma}^{c_2} (\mathbf{x}_2) \rangle \\ 
& = & 2 (\Pi_L)^{c_2 c_1}_{i_2i_1} (\mathbf{x}_2,\mathbf{x}_1) \langle 
\mathbf{\Pi}^{c_1}_{i_1}(\mathbf{x}_1) \rangle \langle \mathbf{
\Pi}^{c_2}_{i_2}(\mathbf{x}_2) \rangle \\ 
3. & & \langle \Delta (\mathbf{\Pi}^a_i(\mathbf{x}) \mathbf{
\Pi}^a_i(\mathbf{x})) 
\bar{\Delta}^{c_1} (\mathbf{x}_1) \bar{\Delta}^{c_2}(\mathbf{x}_2) \rangle
(\Theta^{-1})^{c_1 
d_1}(\mathbf{x}_1, \mathbf{y}_1) (\Theta^{-1})^{c_2 d_2}(\mathbf{x}_2, \mathbf{
y}_2) \langle  
\bar{\Gamma}^{d_1} (\mathbf{y}_1) \rangle \langle \bar{\Gamma}^{d_2}
(\mathbf{y}_2) \rangle \\ & = &  2 (\Pi_L)^{c_2 c_1}_{i_2i_1}
(\mathbf{x}_2,\mathbf{x}_1) \langle
\mathbf{\Pi}^{c_1}_{i_1}(\mathbf{x}_1) \rangle \langle
\mathbf{\Pi}^{c_2}_{i_2}(\mathbf{x}_2) \rangle. 
\end{array}
\nonumber
\eea
Inserting these results into eq.\,(\ref{YM_Kamlah_final1}), we obtain the
results given by
eqs.\,(\ref{Kamlah_overall_corr_electr_ener}-\ref{en_kamlah_2nd_expl})
in the main text. 

\end{document}